\documentclass[paper]{agujournal2019}
\usepackage{url}
\usepackage{lineno}
\usepackage{color}
\usepackage{amssymb,amsmath}
\usepackage{setspace}

\draftfalse

\newcommand{\planss} {{Planatary Space Science }}  
\newcommand{\ssr}{   {Space Sci. Rev. }}

\newcommand{\jgr}{   {J. Geophys. Res.}}
\newcommand{\grl}{   {Geophys. Res. Lett.}}
\newcommand{\apj}{   {Astrophys. J.}}
\newcommand{\apjl}{   {Astrophys. J. Lett.}}

\newcommand{\aap}{   {Astronomy \& Astrophysics}}

\newcommand{\numdataset}{453}
\newcommand{\fullnumdataset}{1261}
\newcommand{\yeardataset}{2022}

\journalname{JGR: Space Physics}

\begin{document}


\title{Thin current sheets in the magnetotail at lunar distances: statistics of ARTEMIS observations}

\authors{S. R. Kamaletdinov\affil{1,2}, A. V. Artemyev \affil{1,2},  A. Runov\affil{1}, V. Angelopoulos\affil{1}}
\affiliation{1}{Department of Earth, Planetary, and Space Sciences, University of California, Los Angeles, USA}
\affiliation{2}{Space Research Institute of Russian Academy of Sciences, Moscow, Russia}

\correspondingauthor{Sergei Kamaletdinov}{sergei2033@g.ucla.edu}

\begin{keypoints}
\item We present a statistical analysis of magnetotail current sheets collected by the ARTEMIS mission for 11 years of observations at $\sim 60$ R$_E$ tail
\item We observe a large population ($\sim 56$\%) of ion-kinetic scale current sheets and a smaller population of partially force-free current sheets ($\sim24$\%)
\item We show that the occurrence rates of intense current sheets and partially force-free current sheets correlate with the solar wind parameters
\end{keypoints}

\begin{abstract}
The magnetotail current sheet's spatial configuration and stability control the onset of magnetic reconnection - the driving process for magnetospheric substorms. The near-Earth current sheet has been thoroughly investigated by numerous missions, whereas the midtail current sheet has not been adequately explored. This is especially the case for the long-term variation of its configuration in response to the solar wind. We present a statistical analysis of 1261 magnetotail current sheet crossings by the Acceleration, Reconnection, Turbulence and Electrodynamics of Moon's Interaction with the Sun (ARTEMIS) mission orbiting the moon ($X \sim -60$ R$_{E}$), collected during the entirety of  Solar Cycle 24. We demonstrate that the magnetotail current sheet typically remains extremely thin, with a characteristic thickness comparable to the thermal ion gyroradius, even at such large distances from Earth’s dipole. We also find that a substantial fraction ($\sim$ one quarter) of the observed current sheets have a partially force-free magnetic field configuration, with a negligible contribution of the thermal pressure and a significant contribution of the magnetic field shear component to the pressure balance. Further, we quantify the impact of the changing solar wind driving conditions on the properties of the midtail around the lunar orbit. During active solar wind driving conditions, we observe an increase in the occurrence rate of thin current sheets, whereas quiet solar wind driving conditions seem to favor the formation of partially force-free current sheets.

\end{abstract}

\section{Introduction}
During the energy-storage phase of magnetospheric substorms (the so-called growth phase), solar wind energy is stored as magnetic energy in the magnetotail North and South lobes that are separated by the near-equatorial magnetotail current sheet (CS). In the subsequent substrom expansion and recovery phases, this energy is transformed into charged particle acceleration and plasma heating due to magnetic field line reconnection in this CS \cite{Baker96,Angelopoulos08}. Observational evidence indicates that during these energy release phases, a significant portion of plasma from the near-Earth plasma sheet is expelled tailward in the form of a plasmoid \cite<e.g.,>{Hoshino94,Angelopoulos13,Li13:plasmoid}. The substorm growth phase is characterized by thinning of the magnetotail CS \cite<e.g.,>{Petrukovich07,Runov21:jastp}, whereas the expansion phase starts with a CS instability largely controlled by the CS's spatial configuration \cite<e.g.,>{Sitnov19}. Therefore, the spatial configuration and dynamics of the magnetotail CS play a key role in the overall magnetospheric dynamics. 

The magnetic field in the magnetotail lobes is in nearly-anti-parallel  directions (towards and away from Earth in the North and South lobes, respectively). Thus, the field configuration enables magnetic reconnection \cite{Coppi66,Vasyliunas75} -- the process which is believed to play an important role in substorm development \cite<see>[and references therein]{book:Gonzalez&Parker}. Considerable effort has been put into theoretical and observational investigation of magnetotail CS \cite<see reviews by>{Baumjohann07,Zelenyi11PPR,Petrukovich15:ssr,Sitnov19,Runov21:jastp}. Analysis of multi-spacecraft missions, like Cluster \cite{Escoubet01}, Time History of Events  and Macroscale Interactions during Substroms \cite<THEMIS, see>{Angelopoulos08:ssr}, and Magnetospheric Multiscale spacecraft \cite<MMS, see>{Burch16} has allowed estimating the typical thickness and spatial profile of the near-Earth magnetotail CS ($X>-30$ R$_{E}$). The near-Earth CS is characterized by current densities of a few $nA/m^{2}$ and consists of a thin CS embedded into a thicker plasma sheet \cite{Asano04:statistics,Runov06,Artemyev11:jgr,Artemyev16:jgr:pressure,Lu19:jgr:cs,Lu20:natcom}. Here, thin CS refers to a current layer with a thickness comparable to the ion gyroradius, which is approximately $\approx 400$ km for a 1 keV ion in the presence of a magnetic field strength of approximately $\approx 10$ nT. This magnetic field strength is commonly observed in the magnetotail lobes at lunar distances \cite<see>{Xu18:artemis_cs,Runov18}.
Observations made by ISEE-3 \cite{Bame83,Slavin83,Slavin85} and  Geotail 
 \cite{Nishida94,Nishida&Ogino98,Nishida00} in the middle ($-50$ R$_E < X <-12$ R$_E$) and distant tail ($X<-50$ R$_E$) revealed the presence of similar thin CSs \cite{Pulkkinen93,Hoshino96,Hoshino97,Vasko15:jgr:cs}. 

From a theoretical perspective, the presence of ion-gyroscale CSs at various distances in the magnetotail, ranging from the near-Earth tail ($X\sim -12R_E$) to the middle and distant tail regions, poses a significant challenge. Such thin current sheets are characterized by strong current, and, thus, necessitate a strong radial gradient in plasma pressure \cite<see>[for discussion of the stress balances in CSs]{Rich72}. However, magnetotail models struggle to incorporate such a strong pressure gradient simultaneously across a wide radial range. Thus, thin CSs either should form locally, within a radially confined region of strong pressure gradient \cite{SB02,Birn04,Liu14:CS}, or should be balanced by alternative mechanisms that do not involve strong pressure gradients \cite<see discussion in>{Artemyev21:grl,An22:currentsheet,Arnold&Sitnov23}. Further statistical investigations of the middle and distant magnetotail CSs should shed light on how thin CSs typically form at these distances. This, in turn, can offer additional insights supporting one of these two scenarios.

Prior CS statistical studies have been based on measurements of near-Earth multi-spacecraft missions. Thus far, information about the midtail CS has been limited to a Geotail statistical survey, covering a short interval of observations \cite<1994–1995, see>{Vasko15:jgr:cs}. 
Due to the lack of spacecraft missions operating within the range of $X\in[-60,-30]$ R$_E$, we adopt the approach proposed by \citeA{Runov18} to compare the properties of the middle tail region (corresponding to lunar distances; $X\sim -60$ R$_E$) with the well-established characteristics of the near-Earth current sheet. 
At such large distances, the influence of the Earth's dipole field is negligible, and the magnetotail CS should be largely controlled by the solar wind properties. 
Indeed, in the magnetotail CS at lunar distances, there is a strong correlation between the ion temperature ($T_{i}$) and the energy of solar wind protons \cite{Artemyev17:jgr:THEMIS&Geotail,Lukin19}.
This suggests that shocked but not yet fully thermalized solar wind plasma plays a crucial role in populating the magnetotail at $X\sim -60$ R$_E$.

In this study we use the measurements made by the two Acceleration, Reconnection, Turbulence and Electrodynamics of Moon's Interaction with the Sun (ARTEMIS) mission probes \cite{Angelopoulos11:ARTEMIS}. 
Originally a part of the THEMIS mission (ThB as P1 and ThC as P2), these identical probes were subsequently repositioned to orbit the Moon, which passes through the magnetotail at $X \sim -60$ R$_E$. This provides us with an opportunity to investigate the properties of the magnetotail at lunar distances.
At such large distances, the CS's flapping motion is rather typical \cite{Sergeev98,Sergeev06}, which can be used to estimate the CS's spatial scale (thicknesses) and current density. ARTEMIS started operating in 2011 and, by now, its measurements encompass 12 years of observations \cite<see large statistics of plasma and magnetic field characteristics in>{Liuzzo22}. In this study, we use data collected between 2012 and 2022, which makes 11 years in total (one full solar cycle). Thus, our CS dataset allows us to reveal how different solar wind conditions impact the properties of the CSs in the magnetotail at lunar distances. In this paper, we aim to address the following two questions: (1) do the characteristics of the near-Earth magnetotail thin CS, e.g, thickness, current density magnitude and embedding differ from characteristics of thin CSs in the the middle tail? (2) To what extent are these characteristics influenced by solar wind conditions? The answer to the first question should provide important data for new CS models \cite<see>{Sitnov&Arnold22,Arnold&Sitnov23,An23:apj}. The answer to the second question should reveal if solar wind is the main driver of thin CS formation in the middle tail.

The paper is organized as follows: In Section \ref{sec:1}, we introduce our observational set consisting of \fullnumdataset$\;$CS crossings and discuss three representative examples. In Section \ref{sec:2}, we describe the selection criteria that enable us to focus on CSs with reliably estimated current density and thickness. Utilizing these criteria, we introduce a subset of \numdataset$\;$CSs from the complete set of \fullnumdataset$\;$CS crossings.
Using this subset, we examine the CS's thickness, current density and force-free character, and discuss relationships between these characteristics. In Section \ref{sec:2:profiles}, we analyze the cross-tail profiles of the main plasma properties surrounding the CSs. In Section \ref{sec:3}, we discuss the relationship of the solar wind parameters on the properties of these magnetotail CSs at lunar distances. In Section \ref{sec:4}, we summarize our results.

\section{ARTEMIS Dataset and Methods}\label{sec:1}
ARTEMIS P1 and P2 are equipped with electrostatic analyzers (ESA) that measure electron ($<30$keV) and ion ($<25$keV) distribution functions which are used to compute both on-board and on the ground plasma moments (ion density $n_i$, ion and electron temperatures $T_i$ and $T_e$) at 4\,s cadence \cite{McFadden08:THEMIS}. Magnetic field data is provided by the Fluxgate Magnetometer (FGM) at a rate of 128 Samples/s with offset stability  $<$ 0.2 nT/12 hr\cite{Auster08:THEMIS}. While most of our analyses rely on P1 measurements, P2 measurements are employed to fill gaps in P1 data. These gaps are likely attributed to operational shadows or resets caused by space weather events like solar energetic particle (SEP) events, particularly in the years 2016 and 2018. We use an aberrated GSM (AGSM) coordinate system, unless otherwise stated. AGSM is introduced by rotating the X-axis of the conventional Geocentric Solar Magnetospheric (GSM) system by approximately 4 degrees. This adjustment accounts for the Earth's relative motion and the disparity between the direction of the solar wind flow and $X_{GSM}$ \cite{Runov05,Vasko15:jgr:cs}. To achieve this, we align the X-axis with the average direction of the solar wind flow. Subsequently, the Y-axis is established by orienting the conventional $Y_{GSM}$ perpendicular to the new X-axis.

We employ the following selection criteria for CS crossings: (1) the spacecraft crosses the neutral sheet ($B_{x} = 0$) and the observed magnitudes of $B_{0}^{+} = max\;B_{x} > 0$ and $B_{0}^{-} = min\;B_{x} < 0$ should be larger than $0.15\; B_{lobe}$, where $B_{lobe}$ is the lobe magnetic field estimated via the vertical pressure balance ($\bold{B}^{2}/2\mu_{0} + n_{i}(T_{e} + T_{i}) \equiv B_{lobe}^2/2\mu_0$) \cite{Baumjohann90}; (2) the vertical pressure balance $B_{lobe}^2/2\mu_0=const$ should be satisfied with $30\%$ accuracy throughout the CS crossing. The first criterion will explore us to sufficiently probe the inner part of the CS, using  $B_{x}/B_{lobe}$ as an effective coordinate to determine cross-tail profiles of plasma properties surrounding the CS. Employing the aforementioned criteria, we manually collected \fullnumdataset~CS crossings in the midtail $X \in [-65,\;-50]$ R$_{E}$ (2012---2022). Using the CS's flapping motion, we convert temporal into spatial variations as explained in the following 2 paragraphs.

To determine the CS's local coordinate system ($\bold{l},\;\bold{m},\;\bold{n}$), we use the Maximum Variance Analysis (MVA)\cite{Sonnerup68}. In the ($\bold{l},\;\bold{m},\;\bold{n}$) system, $\bold{l}$ refers to the direction of maximum variance, $\bold{n}$ refers to the direction minimum variance, and $\bold{m}$ completes the right-handed system. The applicability and accuracy of the MVA rely on the ratios of eigenvalues corresponding to those eigenvector directions, $\lambda_{1}/\lambda_{2}$ and $\lambda_{2}/\lambda_{3}$: the higher these ratios are, the more accurate the estimation of CS's local coordinate system is.  

According to the method developed by \citeA{Sergeev98}, \citeA{Hoshino96} and \citeA{Sergeev06}, one can determine the vertical (along the normal) velocity of the CS flapping motion based on a correlation analysis between the ion bulk velocity component perpendicular to the CS and the time gradient of the maximum variance magnetic field component, $\sim B_x$. We select a subset of \numdataset $\;$ CS crossings where the ion bulk velocity correlates with the magnetic field time gradient (see further discussion for details).

The ARTEMIS mission has covered more than a half of Solar Cycle 24 (2008---2019) and a significant fraction of the ongoing Solar Cycle 25 (2019---today). This temporal coverage provides us with a unique opportunity to investigate magnetotail current sheets under various solar wind conditions. As a result, we can statistically assess the impact of solar wind dynamics on the magnetotail at lunar distances. Towards that goal we employ the Operating Missions as Nodes on the Internet (OMNI) \cite{King&Papitashvili05} dataset, at 1min resolution. This includes the Auroral Electrojet (AE) and the Interplanetary Magnetic Field IMF $\bold{B}$ = IMF ($B_{x},\;B_{y},\;B_{z}$). Subsequently we transform all the IMF components into the AGSM coordinate system. In order to determine this system, we use OMNI data of solar wind speed $\bold{V}_{sw}$. For each CS crossing in our dataset, the values of $AE$ and $V_{SW}$ are prescribed by averaging OMNI 1min data over an interval of 3 hours prior to the crossing. 

\begin{figure}
\centering\includegraphics[width=0.7\textwidth]{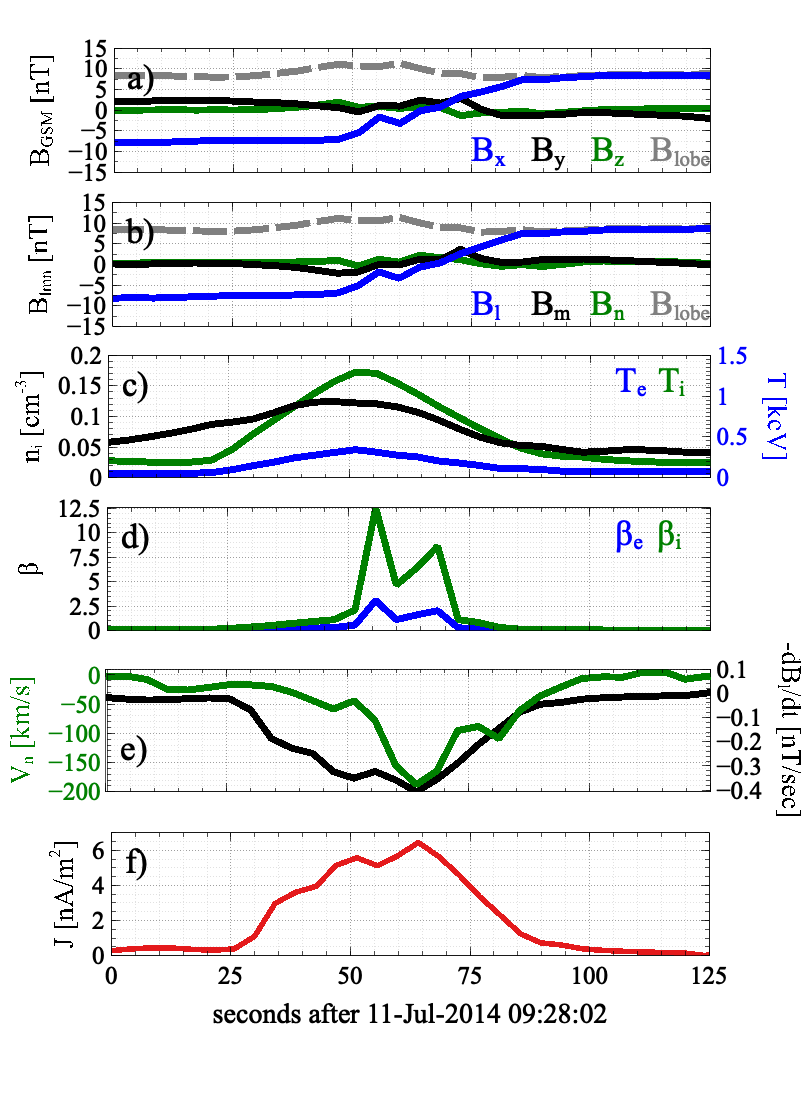}
\caption{The current sheet observed at 2014-Jun-11 by ARTEMIS P1: magnetic field in AGSM coordinates (a), magnetic field in local (MVA) coordinates (b), plasma density, ion and electron temperatures (c), ion and electron $\beta$ (d), ion $V_n$ and $dB_l/dt$ (e), correlation of $V_n$ and $dB_l/dt$ (f), current density as a function of time (g). Main current sheet parameters are: $\lambda_1/\lambda_2= 35.0$, $\lambda_2/\lambda_3=4.0$, $R=0.71$, $L=753$ km, $J_0=8.5$ nA/m$^{2}$
\label{fig1}}
\end{figure}

\begin{figure}
\centering\includegraphics[width=0.7\textwidth]{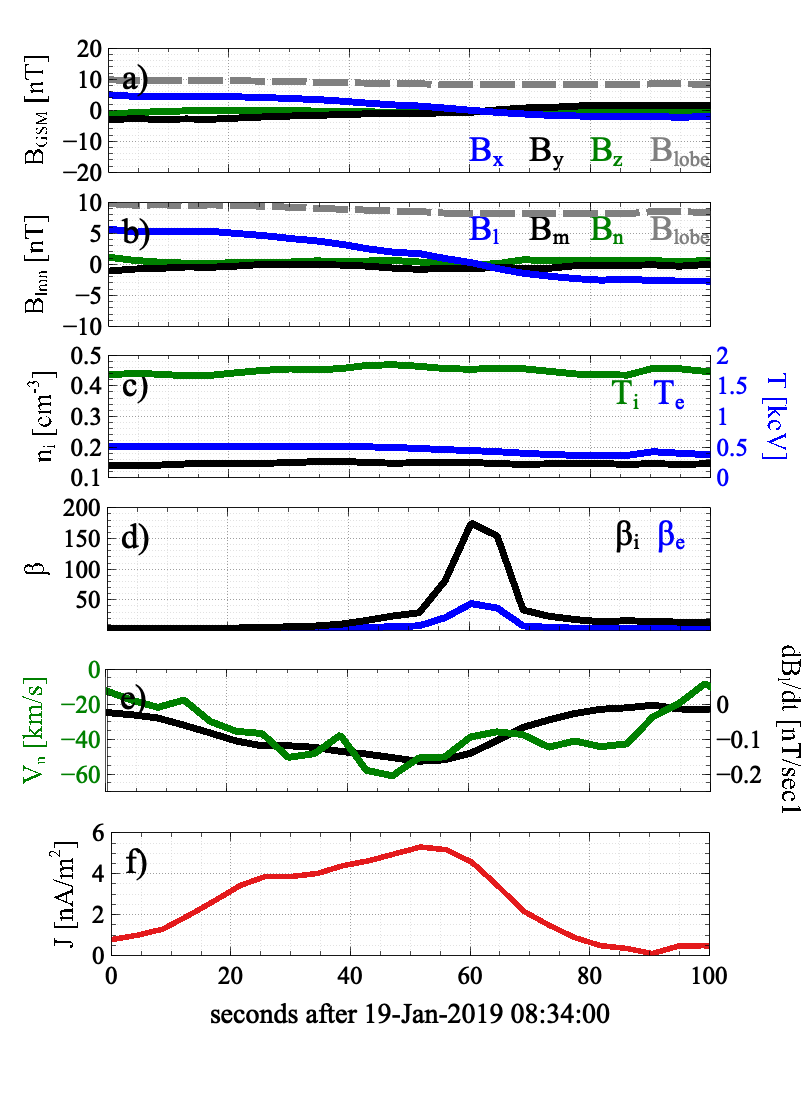}
\caption{The current sheet observed at 2019-Jan-19 by ARTEMIS P1: magnetic field in AGSM coordinates (a), magnetic field in local (MVA) coordinates (b), plasma density, ion and electron temperatures (c), ion and electron $\beta$ (d), ion $V_n$ and $dB_l/dt$ (e), correlation of $V_n$ and $dB_l/dt$ (f), current density as a function of time (g). Main current sheet parameters are: $\lambda_1/\lambda_2=23.3$, $\lambda_2/\lambda_3=11.2$, $R=0.85$, $L=916$ km, $J_0=4.52$ nA/m$^{2}$
\label{fig2}}
\end{figure}

\begin{figure}
\centering\includegraphics[width=0.7\textwidth]{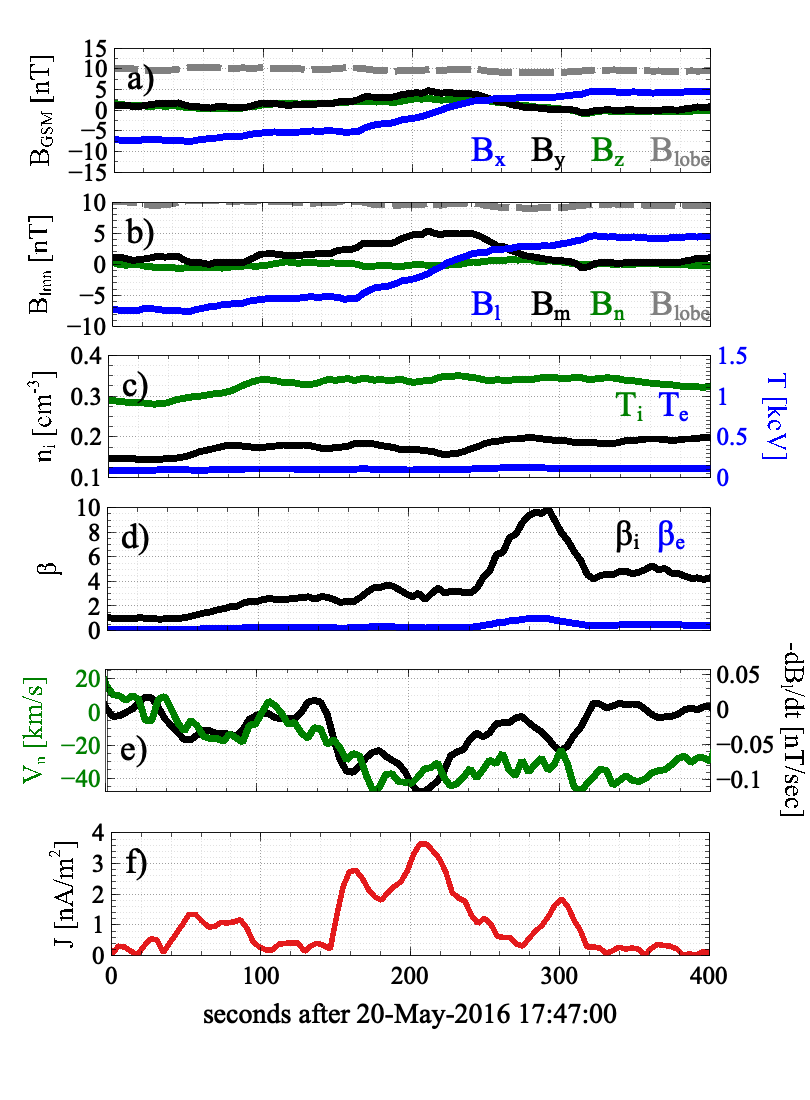}
\caption{The current sheet observed at 2016-May-20 by ARTEMIS P1: magnetic field in AGSM coordinates (a), magnetic field in local (MVA) coordinates (b), plasma density, ion and electron temperatures (c), ion and electron $\beta$ (d), ion $V_n$ and $dB_l/dt$ (e), correlation of $V_n$ and $dB_l/dt$ (f), current density as a function of time (g). Main current sheet parameters are: $\lambda_1/\lambda_2=13.4$, $\lambda_2/\lambda_3=11.0$, $R=0.84$, $L=962$ km, $J_0=5.6$ nA/m$^{2}$
\label{fig3}}
\end{figure}

Figures \ref{fig1}---\ref{fig3} show three examples from the collected CS subset. The CS crossing in Figure \ref{fig1} occurred on 11th July 2014. Figure \ref{fig1}a shows the magnetic field in the AGSM coordinate system. 
A clear reversal of the $B_{x}$ component is observed, resembling a discontinuity, with a relatively symmetric jump from $B_{x} \approx - 7.9$ nT in the southern lobe to $B_{x} \approx +8.5$ nT in the northern lobe.
The shear component $B_{y}$ remains small in comparison to $\Delta B_{x} \approx 16$ nT and the normal component $B_{n}$ fluctuates around zero. We check the pressure balance condition during the entire CS crossing by calculating $B_{lobe} = \left(\bold{B}^{2} + 2\mu_{0}n_{i}(T_{e} + T_{i})\right)^{1/2}$; the resultant $B_{lobe}$ profile is shown by the gray dashed line. $B_{lobe}$ remains almost constant and only slightly varies between $9$--$11$ nT, indicating that pressure balance is satisfied with $10\%$ accuracy. Further, we use the mean value of $B_{lobe}$ as the lobe magnetic field (we use the same notation $B_{lobe}$ for $\langle B_{lobe}\rangle$). Figure \ref{fig1}a presents magnetic field components in the local ($\bold{l},\;\bold{m},\;\bold{n}$) coordinate system. Note that for this CS crossing the local ($\bold{l},\;\bold{m},\;\bold{n}$) coordinate system is well determined, as the ratios of the three eigenvalues are: $\lambda_{1}/\lambda_{2} = 35$ and $\lambda_{2}/\lambda_{3} = 4$, i.e., the three variance  directions are well defined. Figure \ref{fig1}d shows the profiles of ion density $n_{i}$, electron and ion temperatures $T_{e}$ and $T_{i}$. We observe bell-shaped profiles for both density and temperatures. However, the respective maxima are located slightly off the CS's center, $B_{x} = 0$. The ion density varies less than temperature: $n_{i}$ grows from $0.05$ cm$^{-3}$ at the CS's boundaries to $\approx 0.1$ cm$^{-3}$ near the CS's center, whereas $T_{i}$ exhibits a steep increase from $200$ eV at the boundaries to $1.2$ keV near the center. Additionally, the electron temperature $T_{e}$ increases from $60$ eV to around $320$ eV. Comparison of the density and magnetic field profiles shows that this CS crossing resembles a Harris-like CS, where $B_{l} \propto \tanh(z/L)$ and $n_{i} \propto cosh^{-2}(z/L)\propto 1-(B_l/B_{lobe})^2$. However, this Harris-like CS represents a very uncommon event in our dataset (less than a few events). For the overwhelming majority ofthe CSs, the density profiles are observed to be almost flat. Figure \ref{fig1}\,d presents profiles of electron ($\beta_{e}$) and ion ($\beta_{i}$) beta parameters, which are defined as $\beta_{e} = 2\mu_{0}n_{i}T_{e}/B^{2}$ and $\beta_{i} = 2\mu_{0}n_{i}T_{i}/B^{2}$. Both increase towards the CS center, reaching $\beta_{i} \approx 7-13$ and $\beta_{e} \approx 2-3$, respectively. 

We next check the correlation between the ion bulk velocity and the time derivative of the magnetic field in order to estimate the CS vertical speed relative to the spacecraft. In Figure \ref{fig1}e we plot the temporal profile of the normal component of the ion bulk velocity $V_{n}$ and the time derivative of the main magnetic field component, $dB_{l}/dt$. These profiles are correlated, which indicates that the observed crossing was likely caused by the plasma (and magnetic field) motion relative to the spacecraft, i.e., due to the flapping.  We check the applicability of a linear fit by computing thecorrelation coefficient, $R = \left|corr(V_{n}, dB_{l}/dt)\right|$. We find it quite high ($R = 0.71$) which enables us to use the approximation $V_{n} = p_{1}dB_{l}/dt + p_{2}$, where $p_{2}[{\rm km/s}]$ is a velocity offset and $p_{1}[{\rm km/nT}]$ refers to the slope of the fitting. Using $p_{1}[{\rm km/nT}]$ we can estimate the CS's spatial scale as $L_{corr} = 0.5\;|p_{1}|(B_{0}^{+} - B_{0}^{-})$. This method works best when when it is applied to the short sub-interval around the neutral plane, $B_{l} \sim 0$, where the magnetic field spatial profile can be approximated by a linear function $B_{l} \propto z/L$, and thus $dB_{l}/dt \propto V_{n}/L$. 
To enhance the precision of our method, we select a sub-interval centered around the CS that spans approximately 30-60 seconds. We use data only from this sub-interval for the linear fit.

We also estimate the CS's spatial scale by integrating $V_{n}$ over the crossing: $L_{int} = \int_{B = B_{0}^{-}}^{B = B_{0}^{+}} (V_{n} - V_{n0})dt$. Here we take into account possible velocity offset $V_{n0}$ - a steady ion flow which doesn't belong to the flapping motion.  In order to estimate $V_{n0}$, we average $V_{n}$ closer to the lobes, defined as $|B_{l}| > 0.8 B_{0}$ \cite<see similar discussion in>{Sergeev98,Vasko15:jgr:cs}. The two approaches of spatial scale evaluation usually yield results which agree within a factor of $10$. In order to increase the reliability of our estimations, we only use a subset of CS crossings for which 
the discrepancy between these two estimations is within a factor of $4$. For such CSs, we define the thickness $L$ as the mean value of $L_{corr}$ and $L_{int}$. $L$ is then used to estimate the current density peak: $J_{0} = 0.5\;\mu_{0}^{-1}(B_{0}^{+} - B_{0}^{-})/L$. For the CS crossing presented in Figure \ref{fig1}, the linear correlation gives $p_{1} \approx 132$ km/nT and $L_{corr} = 1061$ km. The estimation based on integration gives $L_{int} = 370$ km and thus $L_{corr}/L_{int} \approx 2.37$. This gives $L \approx 753$ km and $J_{0} \approx 8.5$ nA/m$^{2}$. Current density of a few $nA/m^{2}$ represents a typical value for the near-Earth's magnetotail current sheet \cite<e.g.,>{Runov06,Petrukovich15:ssr}. Thickness of a few thousand kilometers, however, indicates a rather thin ion-scale CS \cite<see discussion in>{Runov06,Artemyev11:jgr}. Such thin CSs are believed to be closely associated with plasma instabilities, resulting in magnetic reconnection \cite<see>[and references therein]{Petrukovich16:book,Sitnov19}. We estimate CS's thickness in ion thermal gyroradius $\rho_{i}$, which can be defined either using the maximum magnetic field $B_{0} = max\;|B_{l}|$ at the CS's boundaries \cite<see for example>{Artemyev11:jgr,Vasko15:jgr:cs} or by using the lobe magnetic field $B_{lobe}$ \cite<see>{Runov06}. In this study we use $B_{lobe}$ as a characteristic value of magnetic field strength and define ion thermal gyroradius as $\rho_{i} = m_{i}v_{thi}/eB_{lobe}$, where $v_{thi} = (2\langle T_{i}\rangle/m_{i})^{1/2}$ stands for ion thermal velocity evaluated for the cross-CS averaged temperature. For the CS crossing from Figure \ref{fig1}, we get $\rho_{i}\approx 478$ km, which is comparable to the CS's thickness, $L/\rho_{i} \approx 1.58$. For $\rho_i$ evaluated in $B_0$, we would have $\rho_i\approx 478$ km and $L/\rho_i\approx 1.60$ \cite<see discussion of these two estimates in>{Artemyev11:jgr,Petrukovich15:ssr}. 

Figures \ref{fig2}---\ref{fig3} show two CS crossings from our subset. The CS crossing on January 19 2019 (see Fig. \ref{fig2}) represents the most common type of CS in our set. Figure \ref{fig2}a shows magnetic field profiles in the aberrated GSM coordinate system. The gray dashed line represents $B_{lobe}$ inferred from the pressure balance condition. We see that the magnetic field magnitude at the CS's boundaries, $B_0=(\max B_{l}-\min B_l)/2$, is substantially smaller than the lobe magnetic field $B_{lobe}$. This effect (the so-called CS embedding) is a rather typical property of near-Earth's thin CSs \cite<see>{Runov06}. For the CS from Figure \ref{fig2}, $B_{0}/B_{lobe} \approx 0.25$. Thus, the observed magnetic field strength at the CS's boundary is only $25\%$ of what we should see for a Harris-like configuration. Temporal variations of the plasma properties surrounding the CSs (Fig. \ref{fig2}c) show a relatively flat profile of ion density and temperature, with $n_{i}$ and $T_{i}$ only slightly fluctuating near the constant level of around $0.3$ cm$^{-3}$ and $0.5$ keV, respectively. Therefore, here we see a specific type of CS's spatial configuration in which the spatial distributions of density and temperature are substantially broader than that of the current density, i.e., current density peak is embedded into the plasma density profile. 
Current sheets exhibiting a comparable spatial configuration have been investigated in the near-Earth magnetotail \cite{Runov06,Artemyev11:jgr,Petrukovich15:ssr}. It has been proposed that the embedding of these CSs (i.e., $B_{0}/B_{lobe} < 1$) can be described as an ion-scale thin current sheet enveloped within a thicker sheet, which accounts for density and temperature variations \cite{Sitnov03,Sitnov06,Zelenyi11PPR,Artemyev&Zelenyi13}.

In order to estimate CS's thickness, we compare $V_{n}$ with $dB_{l}/dt$. Figure \ref{fig2}e demonstrates that there is a strong correlation between these two quantities, the  linear correlation coefficient is rather high ($R \approx 0.9$). Applying the same method as previously described, we end up with the following estimations of the spatial scale: $L_{corr} = 801$ km and $L_{int} = 1030$ km. The two methods of estimation agree within a factor of $2$, which allows us to evaluate $L = 915$ km and $J_{0} = 4.5$ nA/m$^{2}$. A small $L$ suggests that this CS is of ion scale, a typical ion gyroradius for this CS is $\rho_{i} \approx 493$ km and $L/\rho_{i} \approx 1.86$. This thin CS differs from the thin CSs observed near the reconnection region for which $B_0/B_{lobe}\sim 1$ \cite{Nakamura06}. However, it does resemble thin current sheets detected in the solar wind, where effective $B_0/B_{lobe}$ varies within a wide range of values \cite<e.g.,>{Vasko21:apjl}. A small $B_0/B_{lobe}$ implies that even a weak plasma pressure variation across the sheet, $\sim \Delta \left(nT_i\right)/nT_i\approx B_0^2/B_{lobe}^2 \ll 1$, is sufficient to balance this CS. Indeed, for this CS, $\Delta \left(nT_i\right)/nT_i = 0.24$ and $B_0^2/B_{lobe}^2 = 0.23$. Such a subtle variation is nearly imperceptible in the density and temperature profiles presented in Fig. \ref{fig2}.

Figure \ref{fig3} shows the third CS crossing. It exemplifies another important type of the CSs that make up a noticeable part of the dataset ($\sim 24\%$). Again, we observe flat profiles of the main plasma properties surrounding the CSs (see Figure \ref{fig3}c). Both density $n_{i} \approx 0.15-0.20$ cm$^{-3}$ and temperature $T_{i} \approx 1.05-1.1$ keV do not exhibit any variations throughout the CS crossing. This results in negligible thermal pressure variation within the CS, $\Delta(n_{i}T_{i})/n_{i}T_{i} < 0.1$ for $B_0/B_{lobe}\approx 1/2$. Despite that, the pressure balance $B_{lobe}\approx const$ is well satisfied to within $2\%$. Thus, we should expect $\Delta \bold{B}^{2} \sim 0$ in order to maintain the full pressure balance $\bold{B}^{2} + 2\mu_{0}n_{i}T_{i} \approx const$ (electron contribution can be omitted due to $T_{e}/T_{i} < 0.1$). According to Figure \ref{fig3}, the main magnetic field component $B_{l}$ varies from $B_{l} \approx -5$ nT to  $B_{l} \approx +5$ nT. We also observe that the normal component remains small through the crossing, $|B_{n}| \lesssim 1$ nT, which is typical of the majority of the CSs in our dataset ($|B_{n}|/B_{0} < 0.3$ for $95\%$ of CSs and $|B_{n}|/B_{0} < 0.1$ for $60\%$ of CSs). The shear component $B_{m}$ exhibits a bell-shaped profile, growing from a zero-value at the CS boundaries up to $5$ nT at the CS's center. Therefore, $B_m$ variation almost precisely compensates for the drop of $|B_{l}|$ and provides $B_{m}^{2} + B_{l}^{2}\approx const$. For 1D current sheets (with only $\partial/\partial r_n\ne 0$) this implies $j_lB_m-j_mB_l={\bf j}\times{\bf B}=0$. Thus, the CSs with such magnetic field configuration are referred to as {\it force-free} CSs. They have previously been reported in the solar wind \cite{Artemyev19:jgr:solarwind,Neukirch20,Lotekar22} and planetary magnetospheres \cite<e.g.,>{Artemyev14:pss,DiBraccio15,Rong15:venus}. This includes individual cases in the Earth's magnetotail at lunar distances \cite{Xu18:artemis_cs} and near-reconnection region \cite{Nakamura08,Artemyev13:jgr}. Theoretical models of such force-free CSs can be found in \citeA{Artemyev11:pop,Mingalev12,Vasko14:angeo_by,Allanson15,Neukirch20:jpp}. In the simplest model of a force-free CS, the spatial profile of the main magnetic field component is identical to that of the Harris sheet, $B_{l} \sim tanh(z/L)$, whereas the shear component profile is $B_{m} \sim cosh^{-1}(z/L)$ \cite{Harrison09:prl}. To quantify the number of similar CSs we introduce $\Delta B_{m}^{2}/B_{0}^{2} = \Delta (B_{m}^{2})/B_{0}^{2}$ parameter, which characterizes the relative contribution of $B_{m}$ into the CS pressure balance ($\Delta B_{m}^2$ is the difference between $B_{m}^2$ value at the CS's boundary and at the center). In a 1D CS, $\Delta (B_{m}^{2})/B_{0}^{2}$ parameter equals to integrated cross-sheet ratio of the two tension force components: $\int{j_lB_m dr_n}/\int{j_mB_l dr_n}$. In the case of a purely force-free current sheet, where $j_lB_m-j_mB_l={\bf j}\times{\bf B}=0$, this parameter equals one. The same parameter has been used in \cite{Xu18:artemis_cs}. For the CS crossing from Figure \ref{fig3}, $\Delta B_{m}^{2}/B_{0}^{2} \approx 1$, a value which corresponds to the purely force-free CS configuration. Statistically, CSs with a pronounced contribution of $B_{m}$ (that is $\Delta B_{m}^{2}/B_{0}^{2} \gtrsim 0.1$) make up about $24\%$ of the events in our set. Note $B_m$ variation with the maximum at $B_l=0$ is generated by local currents $j_l$ and differs well from the conventional flaring 
 $B_m$ component, which drops to zero at $B_l=0$.

\begin{figure}
\centering\includegraphics[width=1.0\textwidth]{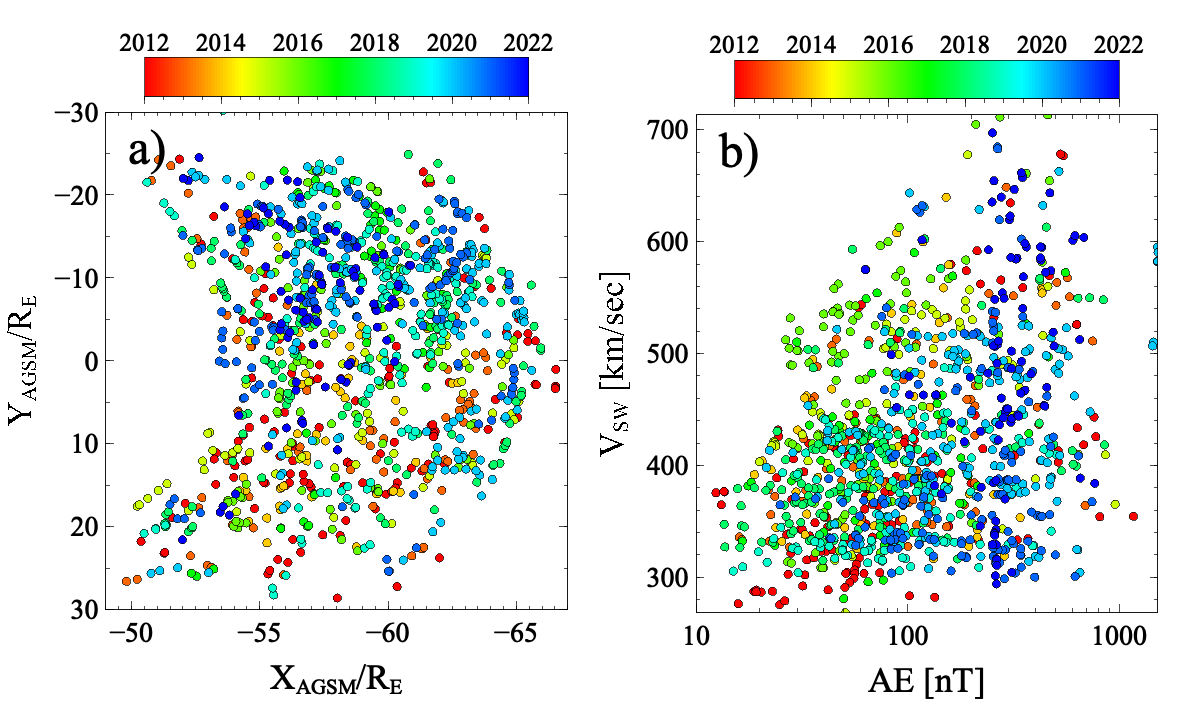}
\caption{Distribution of current sheet crossings in $(X, Y)$ AGSM plane (a), $(AE,V_{sw})$ plane (b), where $AE$ and $V_{SW}$ are averaged over an interval of 3 hours prior to the CS crossing. Color refers to the year of observation. 
\label{fig4}}
\end{figure}

Figure \ref{fig4} presents the distribution of all 1261 CS crossings in our set. The spatial distribution of observed CS crossings on the $X_{AGSM}$---$Y_{AGSM}$(in Earth radii R$_{E}$ units) plane is shown in Figure \ref{fig4}a. The selected set covers $X\in [-65,\; -50]$ R$_{E}$ and $Y\in [-30, 30]$ R$_{E}$, which uniformly resolves both dusk and dawn flanks. Figure \ref{fig4}b shows CSs' distribution in the parametric space of $AE$ and $V_{SW}$. We observe CSs in a wide range of both $AE$ and $V_{SW}$:  $AE$  can vary by two orders of magnitude from $10$ nT to $1000$ nT, whereas CSs are observed during both slow solar wind $V_{SW} \approx 300-400$ km/s and during relatively fast solar wind with $V_{SW}$ reaching up to $800$ km/s. 

\section{Statistics of current sheet characteristics}\label{sec:2}
\begin{figure}
\centering\includegraphics[width=1.0\textwidth]{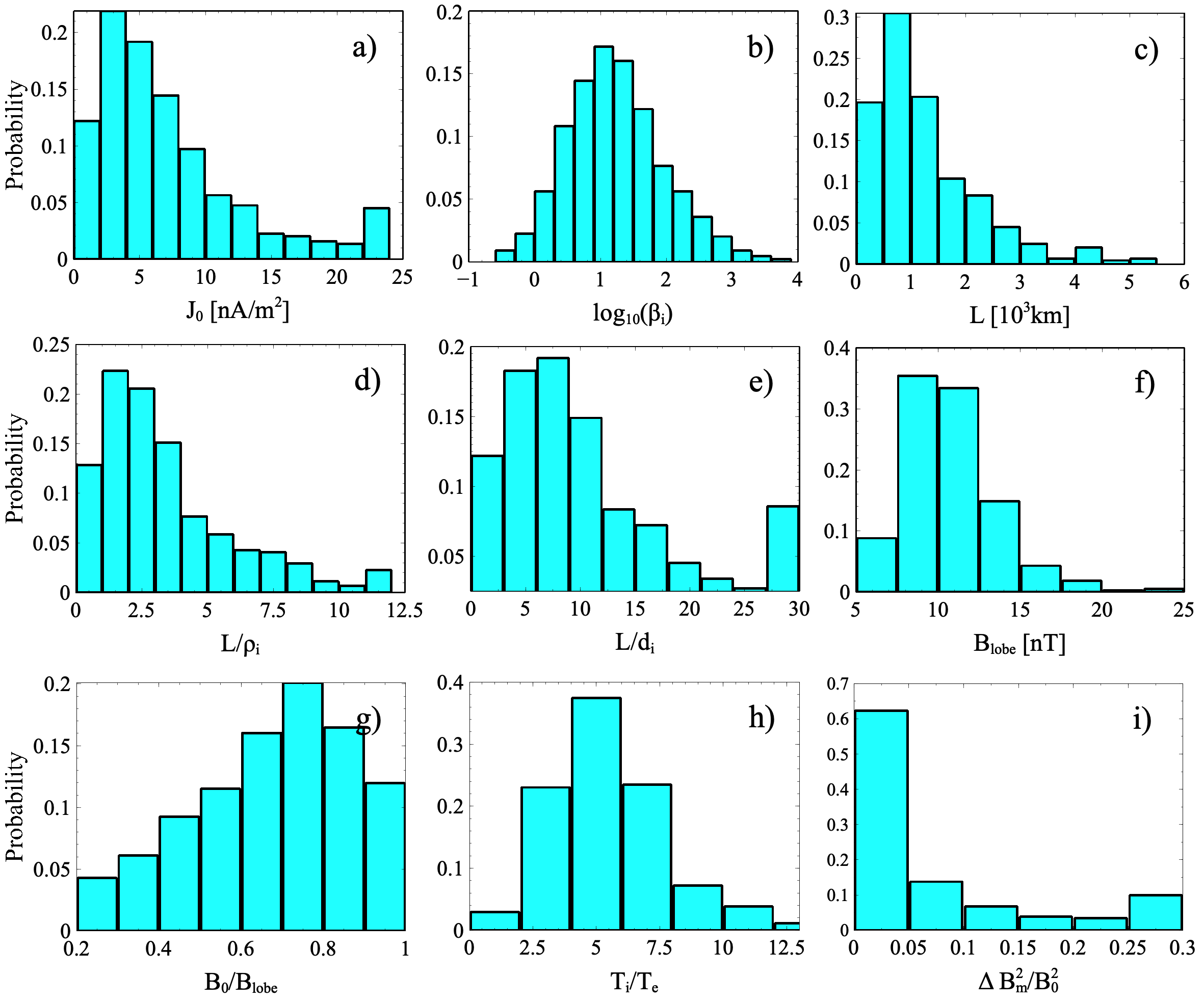}
\caption{Distributions of current sheet characteristics: current density magnitude $J_{0}$ (a),  ion $\beta_{i}$ (b), current sheet thickness $L$ (d) current sheet thickness normalized to ion thermal gyroradius $L/\rho_{i}$, (d) current sheet thickness normalized on ion inertial length $L/d_{i}$ (e), lobe magnetic field magnitude $B_{lobe}$ (f), embedding parameter $B_{0}/B_{lobe}$ (g), ion to electron temperature ratio $T_{i}/T_{e}$, force-free measure $\Delta B_{m}^2/B_{0}^2$  (i)
\label{fig5}}
\end{figure}

Our set consists of \fullnumdataset$\;$CS crossings that cover 11 years of ARTEMIS observations (from 2012 to \yeardataset). Within this set we have a subset of \numdataset$\;$CSs for which we can reliably determine the thickness and current density. For the subset, both linear correlation between $V_{n}$ and $dB_{l}/dt$  and direct integration of $V_{n}$ yield similar (within a factor of $4$) estimations of the spatial scale. Figure \ref{fig5} shows the distribution of the CSs' characteristics from this subset. Typical current density is a few nA/m$^{2}$ (see Figure \ref{fig5}a). These values are similar to those previously observed in the near-Earth magnetotail \cite{Runov06, Artemyev11:jgr, Lu19:jgr:cs} as well as in the distant magnetotail \cite{Vasko15:jgr:cs}. We also notice a substantial population ($24\%$ of the subset) of relatively intense currents ($J_{0} > 10$ nA/m$^{2}$).

Figure \ref{fig5}b presents the distribution of $\beta_{i} = 2\mu_{0}n_{0}T_{i}/B^{2}$ averaged over the CS's center, $|B_l|<0.15B_{lobe}$. It is a log-normal distribution, with the average $\langle \beta_{i} \rangle \approx 100$. Interestingly, the distribution also contains a substantial number of events with $\beta_{i} \lesssim 10$. CSs characterized by such $\beta$ values are not commonly found in the near-Earth magnetotail \cite{Asano03,Asano04:statistics}; however, they are rather ubiquitous in the solar wind environment \cite<e.g.,>{Vasquez07,Artemyev19:jgr:solarwind,Vasko22:apjl}.
In our dataset, more than one third (38\%) of CSs have $\beta_{i} \lesssim 10$. This population is substantially larger than what has previously been observed in the near-Earth magnetotail. This can be attributed to the fact that, compared to the middle and near magnetotail, ion temperature in the magnetotail at lunar distances is usually a few times lower.  

Distribution of CSs' thickness $L$ is shown
in Figure \ref{fig5}c. On average, $L \approx 1000$ km, which is comparable to the ion thermal gyroradius for $1$ keV energy and $\sim 5$ nT of $B_0$ field. Thinner CSs ($L \lesssim 1000$ km) are also observed, comprising 34\% of the entire subset. The number of thick CSs, with $L \gtrsim 10^{4}$ km, is negligible ($< 1\%$). The bias is due to the criteria of CSs' collections: typically flapping velocities don't exceed $V_{N} \approx 10-50$ km/s, whereas most of crossings are fast ($<1$ min). 

Figures \ref{fig5}d-e show the distribution of normalized CS thickness $L$: we use ion gyroradius $\rho_{i}$ and ion inertial length $d_{i}$. Both $\rho_{i}$ and $d_{i}$ are evaluated using plasma parameters at the CS's center; for $\rho_{i}$ we use $B_{lobe}$ as the typical value of magnetic field  strength. The majority of CSs are of ion-gyroscale, the average value is $L/\rho_{i} = 4$ (see Figure \ref{fig5}d), whereas CSs of sub ion-scale ($L/\rho_{i} < 1$) constitute 10\% of the subset. Figure \ref{fig5}f presents the statistical distribution for the lobe magnetic field $B_{lobe}$. 
The average value of $B_{lobe}$ is approximately $10$ nT, which is 2 to 3 times lower than in the near-Earth magnetotail \cite{Runov06,Artemyev11:jgr}. However, it is similar to the observations made by Geotail in the distant magnetotail \cite{Vasko15:jgr:cs}.

On the whole, typical values of $10\;nT$ at $X \in [-60,-50]$ R$_E$ are in a good agreement with predictions of empirical models for the $B_{lobe}$ radial profile \cite{Nakai91,Shukhtina04}. Figure \ref{fig5}g presents the distribution of the embedding ratio $B_{0}/B_{lobe}$, i.e.,  the ratio of magnetic field strength at the CS's boundary $B_{0}$ to the total (lobe) magnetic field $B_{lobe}$. For a Harris-like CS this quantity always equals to one and, therefore, any significant deviation of $B_{0}/B_{lobe}$ from unity would likely indicate an embedded CS configuration. According to Figure \ref{fig5}g, the majority of CSs at lunar distances are weakly embedded. Typically, $B_{0}$ can reach $60\% - 90\%$ of the total magnetic field. Note that the observed $B_{0}/B_{lobe}$ ratio is generally lower in the near-Earth magnetotail, where  $B_{0}/B_{lobe} \sim 0.3-0.4$ \cite<see>{Artemyev08:angeo,Artemyev10:jgr}, i.e., CSs at lunar distances are less embedded and closer to the Harris CS configuration. 

Figure \ref{fig5}h shows the distribution of ion to electron temperature ratio $T_{i}/T_{e}$, computed at the CS's center. In the near Earth's magnetotail CS,for $X \in[-30, -10]$ R$_{E}$, $T_{i}/T_{e}$ typically varies within $[1,5]$ \cite{Wang09,Artemyev16:jgr:pressure,Runov15}. For the CSs at lunar distances, the $T_{i}/T_{e}$ ratio is generally higher than $5$. This indicates that $T_{i}/T_{e}$ decreases earthward and, as a result, electrons should be heated more effectively than ions during earthward transport \cite<see discussion in>{Runov18}. 

Figure \ref{fig5}i presents the analyses of the force-free measure $\Delta B_{m}^{2}/B_{0}^{2}$, where $\Delta B_{m}$ stands for the variance of the shear magnetic field component $B_{m}$ from the CS's center to the boundary. For most of the CSs, the $B_{m}$ variation  remains negligible ($\Delta B_{m}^{2}/B_{0}^{2} < 0.05$ for 65\% of CSs). Nevertheless, 24\% of CSs in our dataset are characterized by a substantial contribution of $B_{m}$ into the pressure balance($\Delta B_{m}^{2}/B_{0}^{2} > 0.1$). These CSs can be classified as partially or even fully force-free \cite<see>{Neukirch20,Neukirch20:jpp}. 

\begin{figure}
\centering\includegraphics[width=1.1\textwidth]{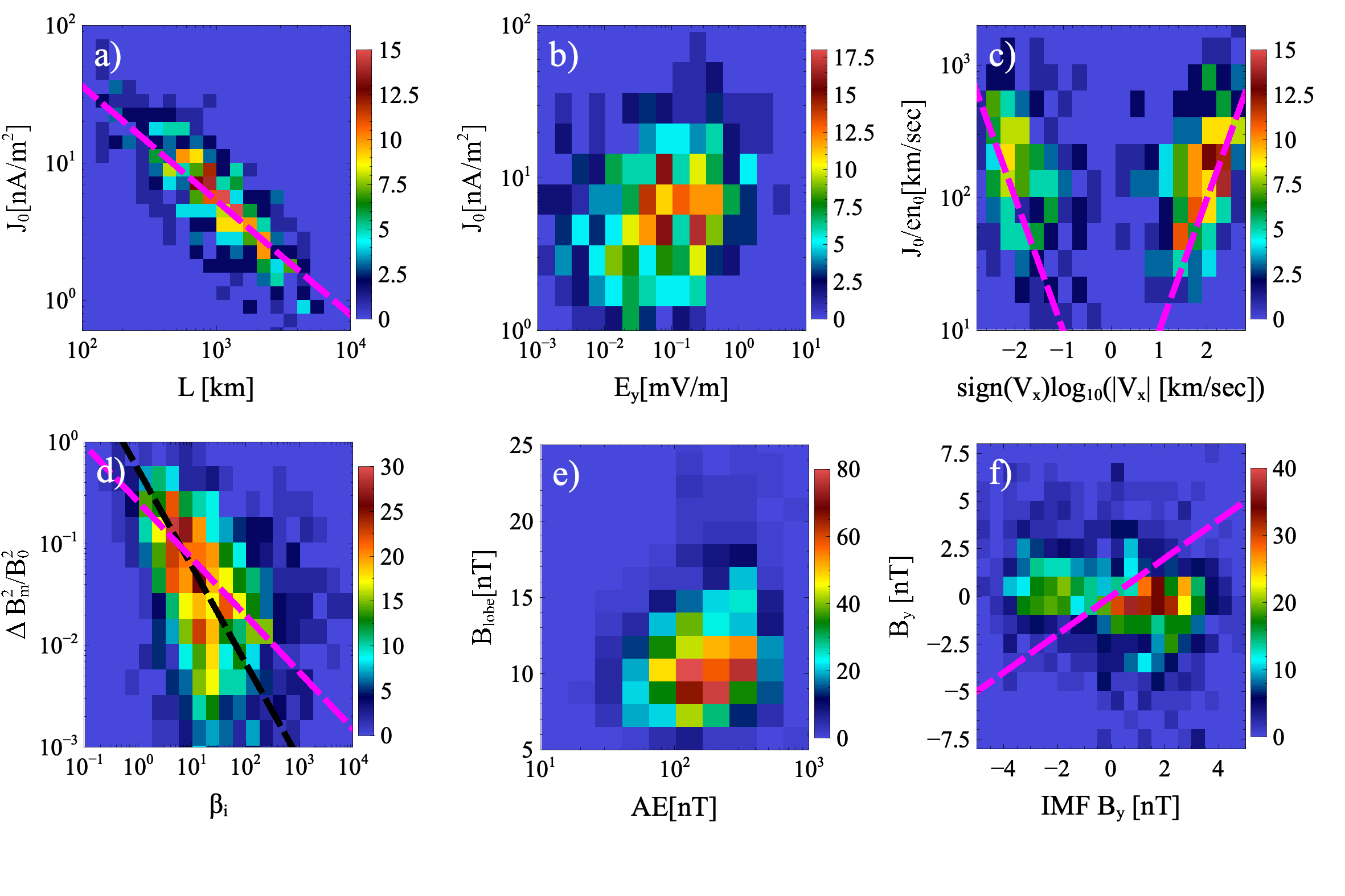}
\caption{Distributions of observed CSs in different parametric spaces: current density magnitude $J_{0}$ versus current sheet thickness $L$, magenta line indicates power law-fit $J[nA/m^{2}] \approx (7.4/L[10^{3}km])^{0.8}$(a), current density magnitude $J_{0}$ versus magnetotail convectional electric field $E_{y}$(b), ion-electron drift velocity $J_{0}/en_{0}$ versus tailward/earthward ion bulk velocity $V_{x}$ , magenta lines show $J_{0}/en_{0} = \pm V_{x}$ (c), force-free measure $\Delta B_{m}^2/B_{0}^2$ versus ion beta $\beta_{i}$, lines show two power-law fits: $\Delta B_{m}^2/B_{0}^2 \approx (0.11/\beta_{i})^{0.55}$(magenta), $\Delta B_{m}^2/B_{0}^2 \approx 0.5/\beta_{i}$(black)(d), lobe magnetic field $B_{lobe}$ versus Auroral Electrojet index $AE$(e), $B_{y}$ component of the magnetic field within the CS versus $IMF B_{y}$, magenta line indicates $B_{y} = IMF\;B_{y}$.
\label{fig6}}
\end{figure}

Using the subset of \numdataset$\;$CSs with reliably determinable current density $J_{0}$ and thickness $L$, we can statistically explore the potential relationship between $J_0$ and $L$ in order to reveal the role of $B_0$ in the CS configuration. There are two main scenarios: (1) $B_0$ is approximately the same for all CSs, and thus $J_0L\approx const$, (2) $B_0$ is proportional to $J_0$ magnitude, while $L$ remains almost the same for all CSs \cite<see discussion in>{Petrukovich11}. Figure \ref{fig6}a presents the distribution of CSs in $J_{0}-L$ space with color bar indicating the number of CSs within each bin. The fitting $J_0\propto 1/L$ describes the observed distribution (the best power-law fitting yields the exponent between -0.9 and -0.8). This indicates that $B_0$ remains constant (actually around a fraction of $B_{lobe}$) for all CSs, which implies that the configuration of thin CSs is primarily governed by the boundary conditions, set by $B_{lobe}$. In this context, the local plasma currents are spatially distributed in such a way that $J_0L\approx const$. Thus, we do not observe any \textit{intense, thick} CSs nor do we observe any \textit{low-intensity, thin} CSs. Note that the near-Earth magnetotail, where $B_0/B_{lobe} \ll 1$, does contain a population of low-intensity, thin CSs \cite< see>{Petrukovich15:ssr}.

Figure \ref{fig6}b shows the number of CSs in $J_{0}-E_{y}$ space, where $E_{y} = |\langle V_{x}B_{z} \rangle |$ is the convection electric field. $V_{x}$ denotes the ion bulk-velocity along $x$-direction, $B_{z}$ is the $z$ component of the magnetic field (in AGSM system), $\langle..\rangle$ denotes averaging over the CS crossing. The distribution of CSs in  $J_{0}-E_{y}$ space indicates that intense CSs (large $J_0$) are usually associated with larger values of convection electric field $E_{y}$. This correlation can be understood within the framework of an embedded CS, supported by a relatively small population of demagnetized {\it Speiser ions} \cite{Speiser65} moving along open trajectories and contributing significantly to the cross-tail current density \cite<see>{Burkhart92TCS,Sitnov00,Zelenyi00}. These ions are effectively accelerated by the convection field $E_y$ \cite{Lyons82,Maha93,Maha94,Zhou11,Zhou12:beamlets}; this acceleration increases the Speiser ion current density. Thin CSs supported by such transient ion populations have been detected in the near-Earth magnetotail \cite{Zhou09,Artemyev10:jgr}. They may, however, be more common in the distant tail, where the equatorial magnetic field is comparatively weaker, allowing for a larger ion population to be demagnetized \cite<see discussion in>{Zelenyi13:UFN}.

To provide an alternative measure of the current intensity, we introduce the ion-electron drift velocity $J_{0}/en_{0}$ ($n_{0}$ -- plasma density). Figure \ref{fig6}\,c provides the number of CS crossings in the  $J_{0}/en_{0} - sign(V_{x})\;\log_{10}(|V_{x}|)$ space. Using $sign(V_{x})\;\log_{10}(|V_{x}|)$ allows us to employ a logarithmic scale while still distinguishing between tailward flows ($V_{x} < 0$) and earthward flows ($V_{x} > 0$). Magenta lines indicate $J_{0}/en_{0} = \pm\;V_{x}$. These results clearly indicate that the most intense CSs occur within fast flows \cite<see also>{Vasko15:jgr:cs}. This could be the result of larger $E_y\propto V_x$ and larger hot ion population, transported by fast plasma flow and supporting the thin CSs. Interestingly, it appears that CSs at the lunar orbit are more often observed within \textit{earthward} flows (for $\sim 60\%$ of events we have $V_{x} > 0$), whereas statistical studies of plasma flows reveal an equal number of tailward and earthward flows \cite{Kiehas18}. 

Typical plasma conditions in the magnetotail at lunar distances (low-$\beta$) seem to favor the formation of partially or even entirely force-free CS configurations. Figure \ref{fig6}d demonstrates that there is an inverse correlation between force-free measure $\Delta B_{m}^{2}/B^{2}_{0}$ and ion beta $\beta_{i}$. The best power-law fitting yields $\Delta B_{m}^{2}/B^{2}_{0} \propto 1/\beta_{i}^{0.55}$(note that we also show $\Delta B_m^2/B_0^2\propto 1/\beta_i$ for comparison). This relationship can be further verified for space plasma systems with a wide range of $\beta$, e.g., Earth's magnetosheath. 

In addition to exploring the distributions of CSs' characteristics, we have also investigated potential connections with solar wind parameters and Auroral Electrojet Index ($AE$). For each CS crossing, we determine  the value of the solar wind speed $\bold{V}_{sw}$, interplanetary magnetic field IMF $\bold{B}$, $AE$ averaged over three hours prior to the crossing. Figure \ref{fig6}e shows that there is a degree of correlation between the lobe magnetic field $B_{lobe}$ and $AE$.  According to the previous studies, $AE$ correlates with the solar wind speed $V_{sw}$ \cite{Hajra2014}. 
Therefore, the observed correlation between $AE$ and $B_{lobe}$ can likely be attributed to the relationship between $B_{lobe}$ and the solar wind dynamic pressure $P_{D} \propto V_{sw}^{2}$, as discussed in \citeA{Nakai91}.

We were unable to find any significant correlation between $B_{lobe}$ and $V_{sw}$ (not shown). 
On the other hand, a similar relationship between the equatorial plasma pressure (approximately $B_{lobe}^2/2\mu_0$) and the $AE$ index has already been reported for the near-Earth magnetotail \cite<see Fig. 5 in>{Baumjohann90}.

Finally, figure \ref{fig6}f shows the distribution of the CSs in the space of IMF $B_{y}$ and CS $B_{y}$: IMF $B_{y}$ stands for $Y$ component of the interplanetary magnetic field, whereas $B_{y}$ is computed at the CS's center. The magenta line indicates IMF $B_{y} = B_{y}$. Note that the ultimate origin of  $B_{y}$ component in the Earth's magnetotail is still unclear. A number of studies have speculated that, at least to some extent, the origin of $B_{y}$ can be traced down to the penetration of the IMF $B_{y}$ to the Earth's magnetosphere \cite{Nakamura08,Petrukovich&Lukin18}. Other studies, however, have suggested that this mechanism for penetration may not be effective in the magnetotail at lunar distances \cite{Xu18:artemis_cs}. Indeed, Figure \ref{fig6}f shows that there is no significant correlation between IMF $B_{y}$ and $B_{y}$.

\section{Profiles of current sheet characteristics}\label{sec:2:profiles}
An essential aspect of the CS spatial configuration is the relationship between spatial (along normal to the neutral plan) distributions of plasma characteristics and current density \cite<see results and discussion for the near-Earth magnetotail CSs in>{Runov06,Artemyev11:jgr,Lu19:jgr:cs}. To determine if a CS is embedded (i.e., CS configuration deviates from a single-component Harris CS), we compare the current density and plasma density profiles.
In the case of an embedded thin current sheet (CS), there are typically two distinct characteristic scales that define its structure. The larger scale is associated with the variations in density and temperature profiles across the CS. Meanwhile, the smaller scale is associated with a intense peak in the current density around the CS's center. 

\begin{figure}
\centering\includegraphics[width=1.0\textwidth]{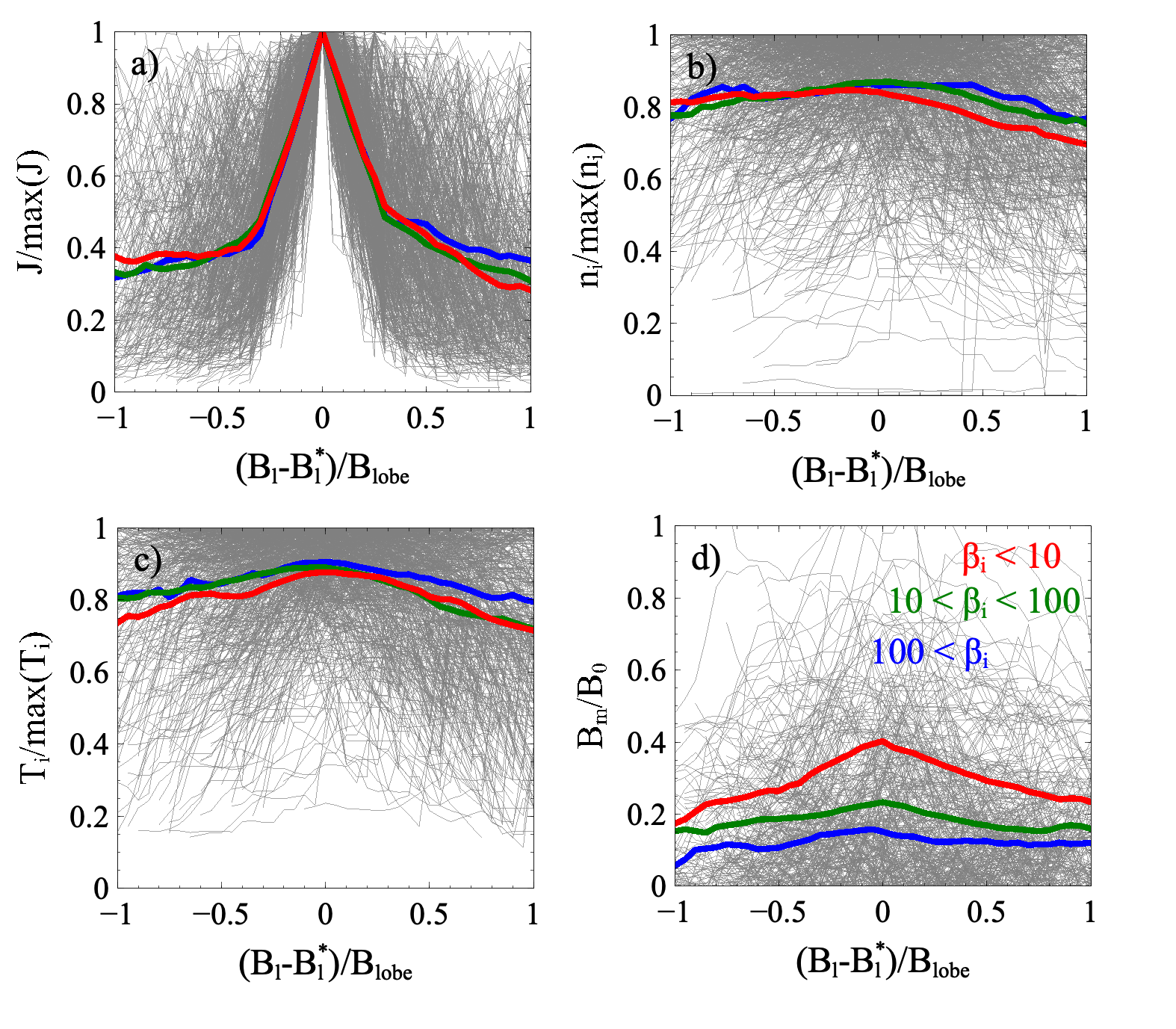}
\caption{Averaged profiles of current sheet characteristics versus $B_{l}/B_{0}$ for different $\beta_{i}$ ranges: current density $J$ normalized on $max(J)$ (a), plasma density $n_{i}$ normalized on it's peak value (b), ion temperature $T_{i}$ normalized on it's peak value (c), shear magnetic field component $B_{m}$ normalized to the magnetic field at the CS's boundary $B_{0}$ (d). 
\label{fig7}}
\end{figure}

We use $B_{l}$ component of the magnetic field (normalized to the lobe magnetic field $B_{lobe}$) as a proxy for the distance to the CS neutral plane, and used $B_l/B_{lobe}$ to shape spatial profiles of plasma characteristics across CSs. Figures \ref{fig7}(a,b,c) show profiles of current density $J$, ion density $n_{i}$, and ion temperature $T_{i}$. For each CS we normalize these profiles to their maximum values and centre the profile at the position of the maximum current density, $B_l-B_l^*$ with $\max J = J(B_{l}^{*})$.  Comparing Figures \ref{fig7}a-\ref{fig7}c reveals that the majority of CSs exhibit a thin current density peak (with a typical thickness of $0.5\;B_{lobe}$), embedded into thicker density and temperature profiles (covering an entire range of $0 < |B_{l}| < B_{lobe}$). Similar to the results for the near-Earth CSs, the ion temperature varies more substantially across the CS compared to the plasma density \cite<see discussion in>{Artemyev17:jgr:THEMIS&Geotail}. 

Figure \ref{fig7}d presents profiles of the shear magnetic field component $B_{m}(B_{l})$, with colors indicating  averaging over CSs in different $\beta_{i}$ ranges. As $\beta_{i}$ decreases, the relative contribution of $B_{m}$ component increases and the $B_{m}(B_{l})$ profile more closely resembles a bell-shaped curve. For CSs with $\beta_{i} < 10$, the relative contribution of $B_{m}$ can reach up to 40\% of $B_{0}$. A magnetic field configuration with a shear component that has a bell-shaped profile, is typically attributed to a force-free equilibrium. There are series of theoretical models describing such an equilibrium for CSs with $B_n=0$ \cite<e.g.,>[and references therein]{Neukirch20,Neukirch20:jpp} and with a finite $B_n\ne 0$ \cite<e.g.,>{Artemyev11:pop,Vasko14:angeo_by,An23:apj}. 

\begin{figure}
\centering\includegraphics[width=1.0\textwidth]{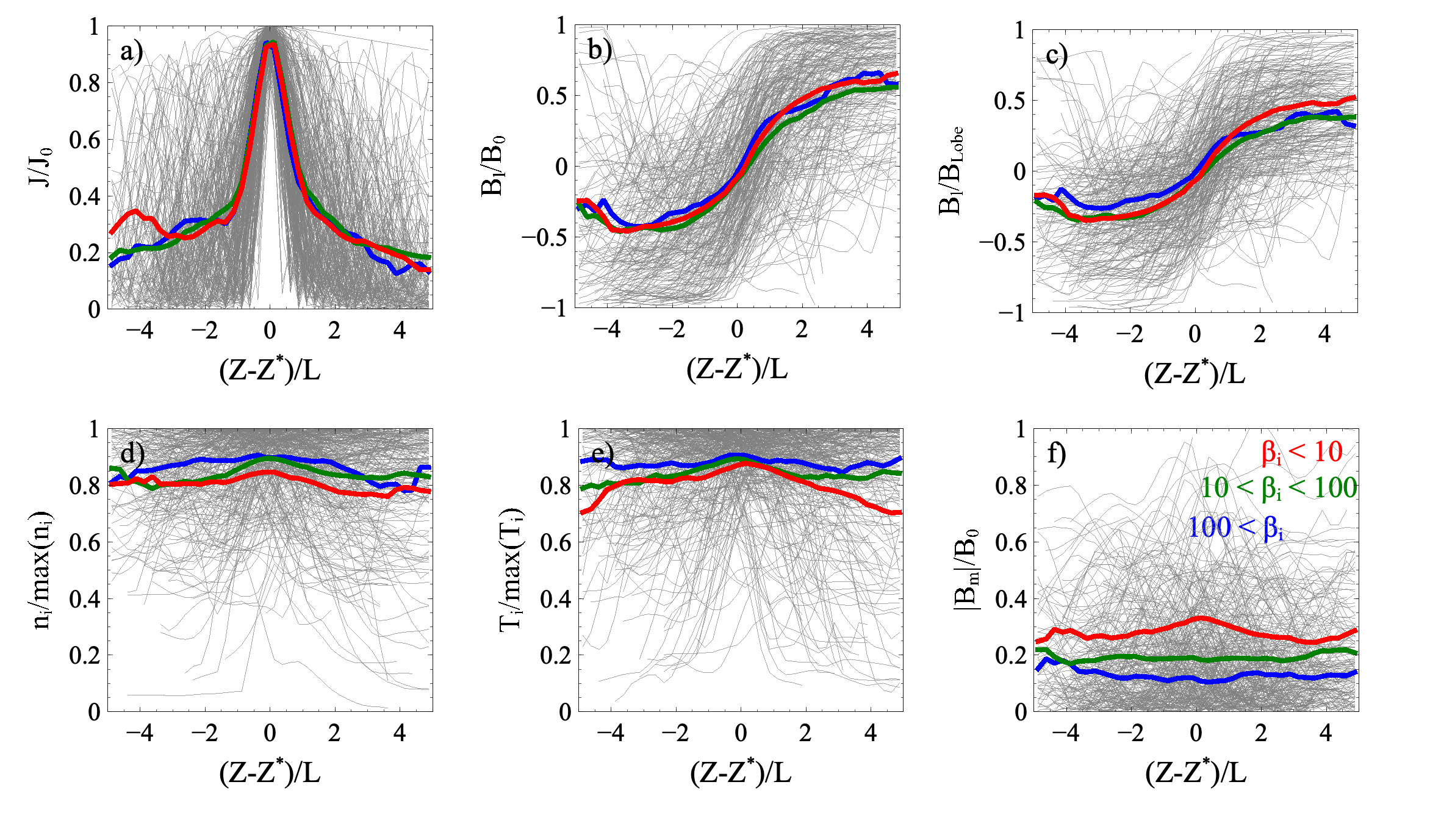}
\caption{Averaged profiles of current sheet characteristics in $Z/L$ space for different $\beta_{i}$ ranges: current density $J$ normalized to the peak value (a), main magnetic field component $B_{l}$ normalized to the magnetic field magnitude at the CS's boundary $B_{0}$ (b), main magnetic field component $B_{l}$ normalized to the lobe magnetic field $B_{lobe}$ (c), plasma density $n_{i}$ normalized in peak value (d), ion temperature $T_{i}$ normalized in peak value (e), normalized shear magnetic field $B_m/B_0$ (f). 
\label{fig8}}
\end{figure}

Figure \ref{fig8} reproduces the profiles previously shown in Figure \ref{fig7}, but in this case, the spatial coordinate along the normal to the current sheet, denoted as $Z(t)$, is used: $Z(t) = \int_{t_{0}}^{t}V_{N}dt$ with $t_{0}$ indicating the time when $B_{l} = 0$. We use $Z-Z^{*}$ in order to account for the offset of the current density peak location $Z^*$ relative to the $B_l=0$. We also normalize $Z-Z^*$ to the CS thickness $L$. Figure \ref{fig8}a presents profiles of $J(Z)$, normalized to the current density peak value for each CS. A typical scale of the central current density peak is $\sim L$, however, a thin CS (thin current layer) is embedded into a larger scale (thicker) sheet. The intensity of this sheet is usually lower ($J \sim 0.3\;J_{0}$) and its typical thickness is $\sim 3-4\;L$. These double-scale current density profiles have also been observed in the near-Earth magnetotail \cite{Petrukovich11,Zelenyi10UFN}. 

Figures \ref{fig8}\,b and \ref{fig8}\,c show spatial profiles of the $B_{l}$ component normalized either to its maximum value $B_{0}$ (Fig. \ref{fig8}(b)) or to the lobe magnetic field $B_{lobe}$ (Fig. \ref{fig8}(c)). For this figure, we orient each $B_{l}$ profile in such a way that $B_{l} < 0$ for $Z-Z^*<0$ and $B_{l} > 0$  for $Z-Z^*>0$. Comparison between Figs. \ref{fig8}b and \ref{fig8}c shows that $B_0$ is quite close to $B_{lobe}$, and the normalized CS configuration does not vary with plasma $\beta$.

Profiles of ion density $n_{i}$ and temperate $T_{i}$ are shown in Figures \ref{fig8}\,d and \ref{fig8}\,e, respectively. 
Typical spatial scales of density and temperature variations are significantly larger than $L$ and thus cannot be directly associated with the observed thin current density peak. Such broad density and temperature profiles are often detected in the near-Earth magnetotail, and represent a typical attribute of the embedded thin CS configuration \cite{Runov06,Artemyev11:jgr,Lu19:jgr:cs}.

Figure \ref{fig8}\,f shows profiles of the normalized shear magnetic field component $B_{m}/B_{0}$. For smaller $\beta_{i}$ values, there is a significant contribution from $B_{m}$ component into the total pressure balance. Interestingly, the $B_m/B_0$ peak scales with the current density peak, but does not show a large-scale variation. Thus, $B_m$ may contribute to the pressure balance in the thin CS, whereas plasma pressure is responsible for the balance of thick embedded CS. Note that a similar spatial configuration is observed for force-free CSs in the solar wind, where $B_m$ profile does not have a small-scale peak, but rather envelopes a thin current layer \cite<see>{Artemyev19:jgr:solarwind}.

\begin{figure}
\centering\includegraphics[width=1.0\textwidth]{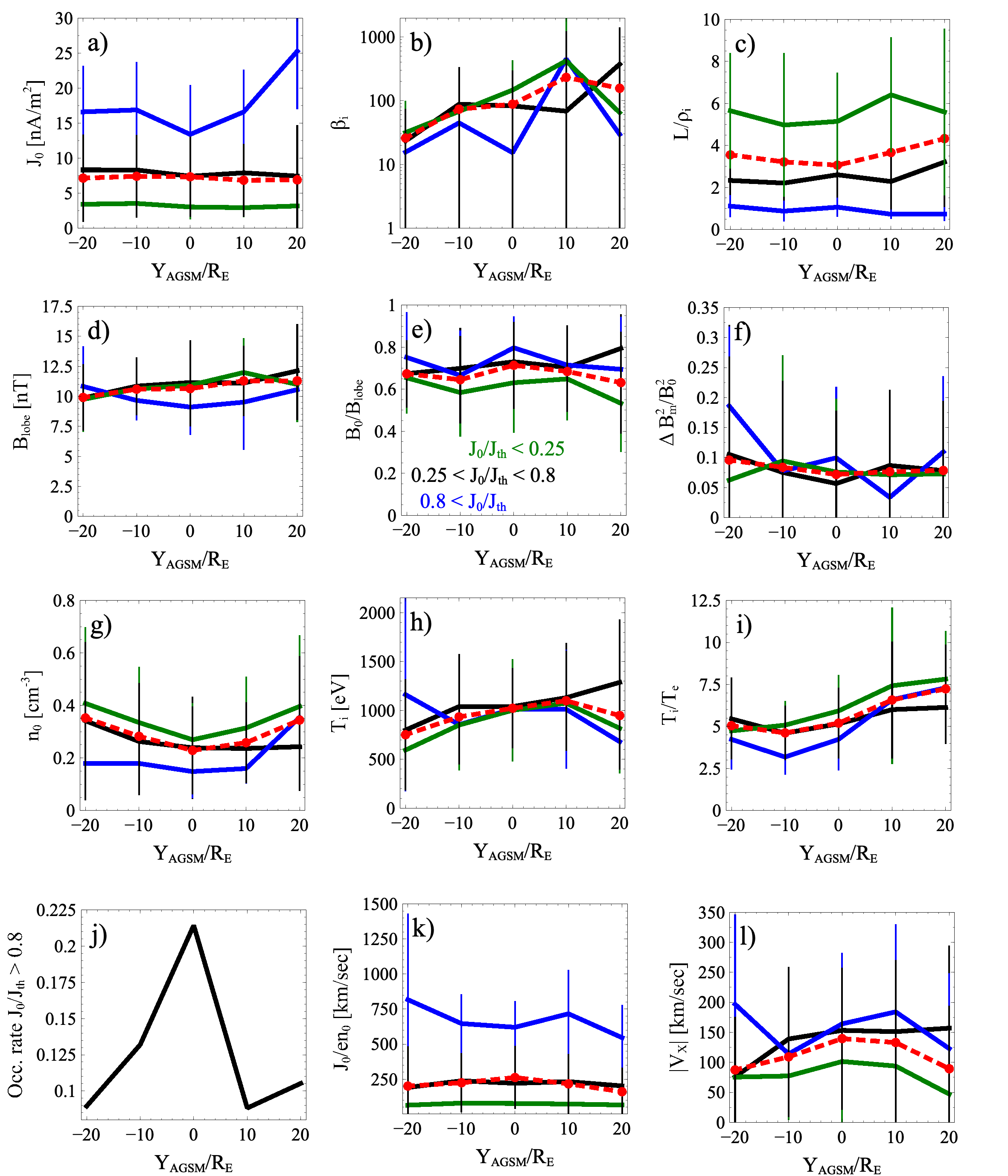}
\caption{Dawn-dusk distribution of current sheet characteristics in different current density ranges, low-intensity $J_{0}/J_{th} < 0.25$(green), intermediate-intensity $0.25 < J_{0}/J_{th} < 0.8$(black), intense $J_{0}/J_{th} > 0.8$(blue), and the average profile for the entire data set (dashed-red): 
magnitude of the current density $J_{0}$ (a), ion-beta $\beta_{i}$ (b), current sheet thickness normalized to ion thermal gyroradius $L/\rho_{i}$ (c), magnetic field strength at the lobes $B_{lobe}$ (d), embedding parameter $B_{0}/B_{lobe}$ (e), force-free measure $\Delta B_{m}^2/B_{0}^2$ (f), ion density $n_{i}$ (g), ion temperature $T_{i}$ (h), ion to electron temperature ratio $T_{i}/T_{e}$ (i), occurrence rate of intense current sheets (j), ion-electron drift velocity $J_{0}/en_{0}$ (k), tailward/earthward ion bulk velocity (l).

\label{fig9}}
\end{figure}

Figure \ref{fig9} presents dawn-dusk distributions of CS characteristics. We use the same AGSM coordinate system defined above. We employ five $10$ R$_{E}$ wide bins, centered at $Y = -20, -10, 0, 10 ,20$ R$_E$. We introduce normalized current density $J_{0}/J_{th}$, where $J_{th} = en_{0}v_{th}$ ($e$ is the elementary charge, $n_{0}$ is ion density in the CS center, $v_{th}$ is ion thermal velocity) and use $J_{0}/J_{th}$ to divide our subset of \numdataset$\;$CSs into 3 subgroups depending on the relative current intensity: (1) low-intensity CSs with $J_{0}/J_{th} < 0.25$, (2) intermediate-intense CSs with $ 0.25 < J_{0}/J_{th} < 0.8$ and (3) intense CSs with $J_{0}/J_{th} > 0.8$. Based on these parameters, low and intermediate intense CSs accounted for $\approx 45\%$ of our subset, each ($\approx 90\%$, cumulatively). Intense CSs represent the residual $10\%$ of the subset. Figure \ref{fig9}\,a shows the dawn-dusk distribution of the average current density $J_{0}$, with colors indicating different sub-groups. Low and intermediate intense CSs do not exhibit any noticeable dawn-dusk asymmetry. For intense CSs, the current density $J_{0}$ trends towards larger $J_0$ at the dusk flank, increasing from $10$ nA/m$^{2}$ at the dawn side to $> 20$ nA/m$^{2}$ at the dusk side. Such dawn-dusk asymmetry is a common feature observed in the near-Earth magnetotail \cite{Artemyev11:jgr,Lu16:cs}, where it is attributed to duskward drift of hot ions \cite<see discussion in>{Walsh14}. However, this phenomenon cannot be easily explained at lunar distances, as a significant sukward drift would require a strong $\partial B_z/\partial x$ gradient, absent at such distances. Importantly, a similar duskward growth has also been reported in middle-tail \cite{Vasko15:jgr:cs}. Note that in CSs with $\partial /\partial x \approx 0$, the classical diamagnetic plasma drift, $\sim {\bf B}\times {\nabla (nT_i)}/B^2 \propto B_x (\partial (nT_i)/\partial z)/(B_x^2+B_z^2)$, becomes zero around $B_x=0$ ($B_l=0$). Therefore, this diamagnetic drift cannot account for duskward drift of hot ions in the presence of a small $B_z \ne 0$ \cite<see discussion in>{Zelenyi10GRL,Artemyev&Zelenyi13}.

Figure \ref{fig9}\,b shows the dawn-dusk profile of $\beta_{i}$. For intermediate intense CSs  this profile seems to be relatively symmetric, whereas intense CSs' $\beta_{i}$ grows ($\times10$) towards the dusk flank. This could be a result of ion temperature increase at dawn flank.

Figure \ref{fig9}\,c shows the dawn-dusk distribution of the normalized thickness $L/\rho_{i}$. The thinnest (and most intense) CSs have typical $L/\rho_{i} \sim 1$ and do not show any dawn-dusk asymmetry of thickness. On the average, observed CSs have $L/\rho_{i} \approx 3-4$, with a slightly larger population of thinner CSs around the midnight sector ($Y \sim 0$), where we expect most of tailward plasma flows with hot ions from near-Earth reconnection region (see \citeA{Kiehas18} and the dawn-dusk temperature profiles below). 

Figures \ref{fig9}\,d presents the dawn-dusk distribution of the lobe magnetic field $B_{lobe}$. For low and intermediate intense CSs, there is only a  weak growth of $B_{lobe}$ from $10$ nT at the dusk flank to $11-12$ nT at dawn flank. For the intense CSs, however, there is a local minimum $B_{lobe} \approx 9$ nT in the midnight sector.

Figures \ref{fig9}\,e shows the distribution of $B_{0}/B_{lobe}$, which characterizes CSs' embedding. In all ranges of $J_{0}/J_{th}$, $B_{0}/B_{lobe}$ distributions are relatively symmetric and show only a subtle increase in $B_{0}/B_{lobe}$ around the midnight sector. Interestingly, it appears that larger values of $B_{0}/B_{lobe} \approx 0.8$ are more typical of intense CSs. 

Figure \ref{fig9}\,f shows the average dawn-dusk distributions of the force-free measure $\Delta B_{m}^{2}/B^{2}_{0}$. Observed profiles for intense and intermediate intense CSs exhibit a modest $\Delta B_{m}^{2}/B^{2}_{0}$ increase towards flanks, reaching levels of $\Delta B_{m}^{2}/B^{2}_{0} \approx 0.1-0.2$. However, no such trend is observed in $\Delta B{m}^{2}/B^{2}_{0}$ for low-intensity CSs. This suggests that force-free CSs don't have any strong preferences in terms of dawn-dusk occurrence. 

Figure \ref{fig9}\,g shows the distribution of ion density in the observed CSs. Density profiles are generally symmetric and feature a minimum at the midnight, with a gradual increase towards flanks. This is similar to what has been observed in the Geotail statistics of CSs \cite{Vasko15:jgr:cs}; the flankward density increase is likely caused by magnetosheath plasma penetration across the equatorial magnetopause. Compared with intense and intermediate intense CSs ($n_{0} \approx 0.2\;cm^{-3}$)., low-intensity CSs seem to be denser ($n_{0} \approx 0.3\;-0.4\;cm^{-3}$).

Figure \ref{fig9}\,h presents dawn-dusk profiles of ion temperature $T_{i}$. intermediate-intense CSs are characterized by a steady growth of $T_{i}$, from $0.8$keV at dawn flank to $1.3$ keV at dusk flank. For low-intensity CSs, a similar growing trend is followed by a sudden drop closer to the dusk flank. This drop is also observed for intense CSs, and seems to concur with a density increase (see Fig. \ref{fig9}\,g). The temperature decrease, and the corresponding density growth closer to the flank, should be attributed to the entry of cold magnetosheath plasma. Figure \ref{fig9}\,i shows dawn-dusk distributions of ion to electron temperature ratio $T_{i}/T_{e}$. For all types of CSs, we see a similar pattern: $T_{i}/T_{e}$ increases from around $4-6\;$ at $Y \approx -15$ R$_{E}$ to $7$ at $Y = 20$ R$_{E}$. So far, we do not have a good explanation for the observed dawn-dusk asymmetry of ion temperature and $T_i/T_e$ ratio in the magnetotail at lunar distances, where $\partial B_z/\partial x\approx 0$ (see previous discussion about the effects of the diamagnetic drift). 

Figure \ref{fig9}\,j presents the occurrence rate of intense CSs, with $J_{0}/J_{th} > 0.8$. The highest relative number of such CSs is detected near the midnight sector, where up to $20\%$ of observed CSs have current densities comparable to $J_{th}$. This number gradually decreases to $5-10\%$ toward the flanks.  

Figure \ref{fig9}\,k shows the dawn-dusk distribution of ion-electron drift velocity $J_{0}/en_{0}$. Profiles of $J_{0}/en_{0}$ for low and intermediate intensity CSs do not exhibit any noticeable dawn-dusk asymmetry and $J_{0}/en_{0}$ is relatively constant across the tail ($40$ km/s for low-intensity CSs and $150$ km/s for moderately intense CSs). For intense CSs, the drift velocity profile is slightly asymmetric and exhibits a local maximum at the dawn flank $Y = -20$ R$_{E}$, reaching $700$ km/s. Similar asymmetry is observed for the dawn-dusk profile of the tailward/earthward velocity $|V_{x}|$ collected in intense CSs' crossing (see Fig.\ref{fig9}\,i). Indeed, Figure \ref{fig9}\,l also shows that intense CSs are usually observed within faster flows. In the case of low-intense CSs, the maximum of $|V_{x}|$ is $\approx 100$ km/s, whereas for intense currents this maximum, and $|V_{x}|$ profile itself, shows almost twofold increase to $|V_{x}| \approx 200$ km/s.  This result suggests that CS's intensity is somehow correlated with the ion flow speed \cite<see discussion in>{Maha94,Zelenyi13:UFN}.

\section{Relation between current sheet characteristics and solar wind parameters}\label{sec:3}
\begin{figure}
\centering\includegraphics[width=0.8\textwidth]{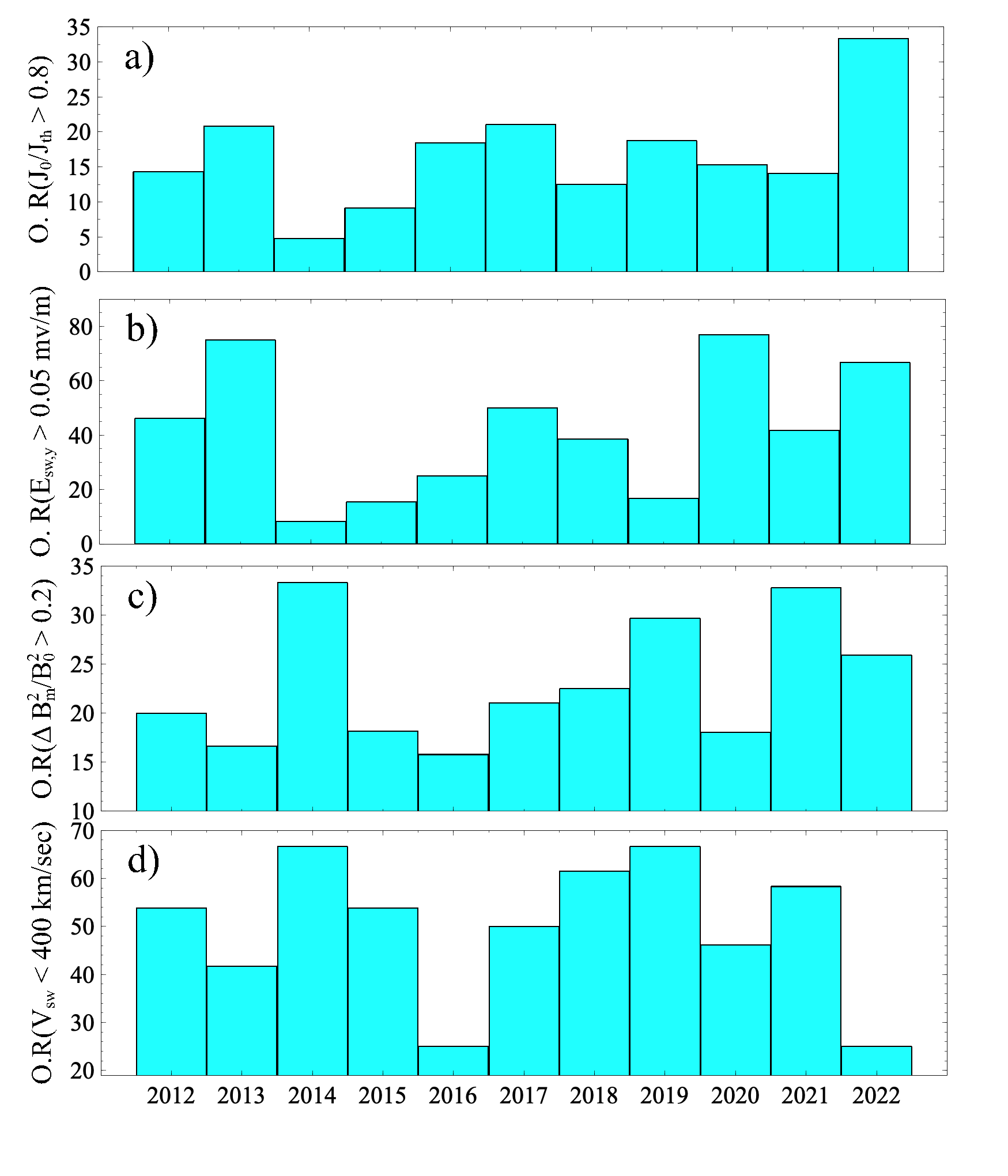}
\caption{Yearly-distribution of CS characteristics and solar wind parameters: 
occurrence rate of intense current sheets $J_{0}/J_{th} > 0.8$ (a), occurrence rate of intense solar wind electric field $E_{sw} = \langle V_{sw,x}*IMF\;B_{z} \rangle > 0.05$ mV/m (b), occurrence rate of CSs with significant contribution of the shear magnetic field component $\Delta B_{m}^2/B_{0}^2 > 0.2$ (c), occurrence rate of slow solar wind  $V_{sw} < 400$ km/s (d).
\label{fig10}}
\end{figure}

Figure \ref{fig10} compares yearly-distributions of averaged CSs' characteristics and solar wind parameters. Figure \ref{fig10}a shows annual occurrence rate of intense CSs, which we define as the relative number of CSs with $J_{0}/J_{th} > 0.8$ within a particular year. Figure \ref{fig10}b presents annual occurrence rate of high values of solar wind electric field $E_{sw,y} > 0.05$ mV/m, where we define $E_{sw,y} = \langle V_{sw,x}\times IMF\;B_{z}\rangle$ with the averaging $\langle....\rangle$ over one full magnetotail crossing by the ARTEMIS mission (a few days long time). Figure \ref{fig10}\,c  shows the occurrence rate of (partially) force-free CSs, i.e., the relative number of CSs with $\Delta B_{m}^2/B_{0}^2 > 0.2$ within a year. Finally, Figure \ref{fig10}\,d presents the occurrence rate of slow solar wind flows $V_{sw} < 400$ km/s. Comparison between Figures \ref{fig10}\,a and \ref{fig10}\,b highlights a similarity in occurrence rates of intense CSs and strong magnetosphere driving. Both occurrence rates maximize in 2012---2013 and undergo a abrupt drop in 2014, which is then followed by a slow increase until 2017. According to Figure \ref{fig10}\,d, this abrupt drop in $E_{sw}$ can be attributed to the decrease in the average solar wind speed. The latter also manifests itself in a higher occurrence rate of slow solar wind flows in 2014 (see Fig. \ref{fig10}\,d). Interestingly, the drop in $E_{sw,y}$ coincides with the start of the {\it early declining phase} of the 24th solar cycle (SC-24) and the abrupt reduction of the geomagnetic activity \cite{Rawat2018SC23_24}. The declining phase of a solar cycle is usually characterized by a growing number of high-speed streams ($V_{sw}\sim 700-800$ km/s) in the solar wind \cite{Bame1976,Gosling1976,Grandin19HSS}, which are thought to have a substantial impact on the geomagnetic activity \cite{Tsurutani2006}.

Recently, \citeA{Grandin19HSS} analyzed the yearly distribution of high-speed streams during solar cycles 22-24 and found that the early declining phase of SC-24 was characterized by an abrupt drop in the number of high-speed streams. This may help us to explain the apparent decrease of the average $V_{sw}$ in 2014 and the subsequent reduction of magnetosphere driving,  $E_{sw,y} \propto V_{sw}$. As the occurrence rate of $E_{sw,y} > 0.05$ mV/m starts to increase in 2015 and 2016, this increase is accompanied by a similar rise in the relative number of intense CSs in the magnetotail at lunar distances. Both occurrence rates reach maxima during the late declining phase in 2017-2018 years. This observation is not surprising and is, in fact, consistent with the idea that the number of high-speed streams reaches its maximum just before the SC-24 minimum \cite{Grandin19HSS}. A pronounced decrease in the occurrence rate of low solar wind flows in 2016 (see Fig. \ref{fig10}d) supports this explanation. Overall, obtained distribution of $E_{sw,y}$ looks similar to the results from \citeA{Rawat2018SC23_24}, in which two local maxima of SC-24 $E_{sw,y}$ were found around 2013 and 2016. 

A substantial number of observed CSs in our dataset can be characterized as partially force-free magnetic field configurations. According to Figure \ref{fig10}c, the yearly-distribution of CSs with $\Delta B_{m}^2/B_{0}^2 > 0.2$ varies from $15\%$ to $35\%$. Comparing Figures \ref{fig10}c and \ref{fig10}d, we notice a weak but rather persistent correlation between the occurrence rate of force-free CSs and the occurrence rate of slow solar wind flows: both rates reach maximum during the yearly declining phase of SC-24 in 2014-2015 and during the transition from SC-24 to SC-25. The source of this correlation remains unclear. One potential explanation could be that the ion temperature $T_{i}$ in the magnetotail at lunar distances correlates with the energy of solar wind protons $m_{p}V_{sw}^2/2$, \cite{Artemyev17:jgr:THEMIS&Geotail, Lukin19}. By invoking this correlation $T_{i} \propto V_{sw}^{2}$, one can speculate that lower values of the average $V_{sw}$ should result in statistically lower $\beta_{i} \propto T_{i}$ in the distant magnetotail. This, in turn, should favor formation of force-free configurations of the magnetic field at lunar distances (see Fig. \ref{fig6}d).

\section{Conclusions}\label{sec:4}
In this study we have presented a statistical analysis of \fullnumdataset$\;$current sheet crossings in the Earth's magnetotail at distances $ \sim 60\;R_{E}$. We use the data provided by the Acceleration, Reconnection, Turbulence and Electrodynamics of Moon’s Interaction with the Sun  mission and combine them with the technique proposed by \citeA{Hoshino96,Sergeev98}, thereby making it possible to reliably estimate current density and thickness for the subset of \numdataset$\;$current sheets. Our main findings pertaining to the properties of these CSs can be summarized as follows:
\begin{enumerate}
  \item  Current densities are typically a few $\sim $ nA/m$^{2}$, which is similar to the results previously reported in the near-Earth magnetotail. However, we do observe a fraction $\sim 23\%$ of intense current sheets with current density $J_{0} > 10$ nA/m$^{2}$
  
  \item Most of the observed current sheets are relatively thin, with a typical thickness $L \sim 10^{3}$ km that is usually about the ion thermal gyroradius $\rho_{i}$. Statistically, the average thickness is approximately a few ion gyroradii  $L/\rho_{i} \sim 1-5$. 
  
  \item The comparison between the ion-electron drift velocity $J_{0}/en_{0}$ and the tailward/earthward flow velocity $V_{x}$ indicates that more intense current sheets are observed within faster tailward/earthward flows. The distribution of $J_{0}/en_{0}$ versus $V_{x}$ exhibits a small earthward asymmetry, which might indicate that some fraction of observed currents is associated with distant tail reconnection. 
   
  \item Spatial profiles of the main plasma properties surrounding the current sheets show that typical current sheets are weakly embedded with $B_0/B_{lobe}\approx 0.8$, which is higher than $B_0/B_{lobe}\approx 0.3-0.5$ reported for the near-Earth magnetotail.
   
   \item  As many as a quarter of the observed current sheets are characterized by a significant variation in the shear magnetic field component $B_{m}$, with the $B_m$ peak at the current sheet center, i.e., this population of current sheets are partially force-free. 
\end{enumerate}

Thanks to the duration of the ARTEMIS mission, we were able to address the question of how different solar wind conditions may influence the properties of the magnetotail current sheet at the lunar distances. Here we report the following findings 
\begin{enumerate}
  \item The occurrence rate of intense current sheets ($J_{0}/J_{th} > 0.8$) is correlated with the occurrence rate of strong magnetosphere driving, $E_{sw,y} = \langle V_{sw,x}*IMF\;B_{z} \rangle > 0.05$ mV/m. By recalling the established inverse relation between the current density and the thickness, we argue that fast solar wind should drive the formation of thin current sheets. 
  \item  The occurrence rate of current sheets with significant variation of $B_{m}$ component (partially force-free current sheets) is correlated with  the occurrence rate of slow solar wind flows.
\end{enumerate}

Previously published results have shown an abundance of ion-gyroscale (thin) CSs in the near-Earth magnetotail. An important implication of this study is the similar substantial population of thin current sheets observed in the middle tail, extending as far as $X \sim - 65$ R$_E$. All of this poses a significant challenge for the current CS models: how to explain the formation of thin CS within the entirety of $x\in[-10,-65]$ R$_E$ domain? The most developed class of theoretical CS models with the dawn-dusk current $\propto \partial (nT_i)/\partial x$ \cite<e.g.,>[and references therein]{SB02,Birn04} cannot explain the formation of such a CS with $L/L_x\approx 1000$ km $/50$ R$_E \ll B_z/B_0$ \cite<see discussion in>{Sitnov&Merkin16,Sitnov19:jgr,Artemyev19:jgr:ions,Artemyev21:grl}. The prospective approach for modeling such a long, thin CS includes incorporating anisotropic and non-gyrotropic terms into the CS pressure balance \cite<see discussion in>{An22:currentsheet,Sitnov&Arnold22,Arnold&Sitnov23}. Therefore, our results confirm the vital need for  non-ideal ion physics in modeling of the magnetotail CS.

\acknowledgments
We acknowledge support by NASA awards 80NSSC20K1788, 80NSSC22K0752, and 80NSSC18K1122. We acknowledge NASA contract NAS5-02099 for use of data from the THEMIS mission. 
We thank K. H. Glassmeier, U. Auster, and W. Baumjohann for the use of FGM data provided under the lead of the Technical University of Braunschweig and with financial support through the German Ministry for Economy and Technology and the German Aerospace Center (DLR) under contract 50 OC 0302. 

\section*{Open Research}
\noindent ARTEMIS data is available at http://themis.ssl.berkeley.edu. Data access and processing was done using SPEDAS V4.1 \citeA{Angelopoulos19} available at \url{https://spedas.org/}.


\end{document}